\documentclass[]{jfm}

\usepackage{graphicx}
\usepackage{epstopdf,epsfig}
\usepackage{newtxtext}
\usepackage{newtxmath}
\usepackage{natbib}
\usepackage{hyperref}
\hypersetup{
    colorlinks = true,
    urlcolor   = blue,
    citecolor  = black,
}

\newcommand{\RomanNumeralCaps}[1]
\usepackage{amsmath, amssymb}
\usepackage{caption, subcaption}
\usepackage{xcolor, ulem}

\newcommand{\bm}[1]{\mbox{\boldmath$#1$\unboldmath}}
\newcommand\Ha{\mbox{\textit{Ha}}}  % Hartmann number
\newcommand{\ie}{i.e.\ }
\newcommand{\cf}{c.f.\ }

\newcommand{\yd}[1]{{#1}}
\newcommand{\mbr}[1]{{#1}}
\newcommand{\tb}[1]{{#1}}

\title{Energy stability of magnetohydrodynamic flow in channels and ducts}

\author{Thomas Boeck\aff{1}
  \corresp{\email{thomas.boeck@tu-ilmenau.de}},
  Mattias Brynjell-Rahkola\aff{1}\corresp{Present address: Department of Applied Mathematics and Theoretical Physics, University of Cambridge, Centre for
Mathematical Sciences, Wilberforce Road, Cambridge CB3 0WA, United Kingdom}
 \and Yohann Duguet\aff{2}}

\affiliation{
  \aff{1}Institute of Thermodynamics and Fluid Mechanics, Technische Universit\"at Ilmenau, P.~O.~Box 100565, 98684 Ilmenau, Germany
  \aff{2}LISN-CNRS, Universit{\'e} Paris Saclay, F-91400 Orsay, France
}

\begin{document}
\maketitle

\begin{abstract}
  We study the energy stability of pressure-driven laminar magnetohydrodynamic flow in a rectangular duct with transverse homogeneous magnetic field and electrically insulating walls. For sufficiently strong fields, the laminar velocity distribution has a uniform core and convex Hartmann and Shercliff boundary layers on the walls perpendicular and parallel to the magnetic field. The problem is discretized by a double expansion in Chebyshev polynomials in the cross-stream coordinates. The linear eigenvalue problem for the critical Reynolds number depends on the streamwise wavenumber, Hartmann number and the aspect ratio. We consider the limits of small and large aspect ratios in order to compare with stability models based on one-dimensional base flows. For large aspect ratios we find good numerical agreement with results based on the quasi-two-dimensional approximation. The lift-up mechanism dominates in the limit of zero streamwise wavenumber and provides a linear dependence between the critical Reynolds and Hartmann number in the duct.  The duct results for small aspect ratio converge to Orr's original energy stability result for  spanwise uniform perturbations imposed on the plane Poiseuille base flow. We also examine  different possible symmetries of eigenmodes  as well as the purely hydrodynamic case in the duct geometry.
\end{abstract}

\begin{keywords}
%  Authors should not enter keywords on the manuscript, as these must be chosen by the author during the online submission process and will then be added during the typesetting process (see \href{https://www.cambridge.org/core/journals/journal-of-fluid-mechanics/information/list-of-keywords}{Keyword PDF} for the full list).  Other classifications will be added at the same time.
\end{keywords}

%{\bf MSC Codes }  {\it(Optional)} Please enter your MSC Codes here

\section{Introduction}
\label{sec:introduction}

The main goal of hydrodynamic stability theory is to predict the parameters for which a given laminar flow can lose its stability and, possibly, turn turbulent. It requires monitoring both the short-time and the long-time  fate of infinitesimal disturbances to the so-called base flow \citep{schmid2001stability}. The concept of energy stability threshold is a key element of the associated toolbox. It refers to the largest value of the governing parameter (here the Reynolds number) below which the kinetic energy of \textit{all} disturbances decays monotonically in time, regardless of their amplitude. For many academical flow cases the value of that threshold, denoted $\Rey_E$, matches exactly the value above which unstable modes are found. For flows characterised by strong non-normality of the associated linear operator, however, $\Rey_E$ lies strictly below the onset of instability. Such flows include most incompressible flows dominated by shear. Rather than separating stable from unstable regimes, it divides the real $\Rey$-axis into a lower range ($\Rey \le \Rey_E$) where all disturbances monotonically decay, and an upper range  ($\Rey > \Rey_E$) where energy growth is momentarily possible, possibly transient, at least for well-chosen initial conditions. Early historical examples of energy stability calculations include the works of Joseph, Busse and co-authors in simple subcritical flow configurations such as plane Couette flow, plane Poiseuille flow or pipe flow \citep{joseph1969stability,joseph1971place,busse1969bounds,busse1972property}, which have been revised recently \citep{falsaperla2019nonlinear,xiong2019conjecture,nagy2022enstrophy}. The transition to turbulence in such flows is known to be subcritical in Reynolds number, and to be dominated by linear yet non-normal effects \citep{trefethen1993hydrodynamic,reddy1993energy}.
Since the exact transition thresholds for actual subcritical transition is statistical it is typically difficult to evaluate \citep{lemoult2016directed,kashyap2022linear}.
The value of $\Rey_E$ given by energy stability theory appears as a much simpler quantity to evaluate in practice, since it is based mostly on linear mechanisms and is perfectly well-defined mathematically speaking.

Energy stability remains an important robustness indicator also for stable flow regimes, as it indicates a \textit{safe} range of Reynolds numbers in which the flow can be operated without any risk of transition. Additional forces acting on a given flow affect the momentum and the energy balance, which can have a quantitative  repercussion on the value of $\Rey_E$. We focus in this article on flows of liquid metals in channels and ducts in the presence of an imposed magnetic field.
While this configuration is relevant for certain applications such as liquid metal cooling systems for fusion reactors \citep{muller2001magnetofluiddynamics}, it remains a simplified configuration that is of fundamental interest in magnetohydrodynamic research since the beginning of the field \citep{hartmann1937paper}.
For the parameters under study, the magnetic Reynolds number $\Rey_m$ is small enough so that the classical low-$\Rey_m$ approximation \citep{muller2001magnetofluiddynamics} holds, and no induction equation needs be taken into account. The magnetic field, depending on its orientation,  generates Lorentz forces inside the flow which can modify the net force balance, while the presence of an electrical current contributes to increased dissipation. In particular, the global stability of the laminar flow can be enhanced if all velocity perturbations are damped by magnetic effects. This results in transition being delayed to higher Reynolds numbers, a property easily quantified by monitoring $\Rey_E$ (although the value of $\Rey_E$ underestimates in this case the exact values of $\Rey$ where transition occurs).

Specifically, the magnetohydrodynamic (MHD) duct accommodates two different types of boundary layers, namely the Hartmann and the Shercliff layers \citep{knaepen2008magnetohydrodynamic}.
%{\color{cyan}MBR: Since we are citing Mueller \& Buehler heavily in the preceding and succeeding sentences, it appears strange to suddenly cite Knaepen \& Moreau for this basic claim. Maybe we should stick to Mueller \& Buehler unless you really want to cite Knaepen \& Moreau in which case we might add them in more places.}\yd{We are citing KM's review as an alternative to  MB's book.}
These are respectively orthogonal and parallel to the applied magnetic field.
For a unidirectional fluid flow subject to an externally imposed magnetic field, the interaction between the fluid motion and the magnetic field imposes a difference in electric potential between the Shercliff walls that drives a transversal electric current density.
Assuming that the walls are electrically insulating, conservation of charge makes this current turn and reverse through the Hartmann layers such that closed current streamlines are formed.
Due to such a reversal in the flow of charges, the Lorentz force, which is proportional to the current, tends to impede the fluid motion in the bulk and simultaneously accelerate the flow within the Hartmann layers \citep{muller2001magnetofluiddynamics}.
This in turn leads to Hartmann and Shercliff layers with different thicknesses: for the former it is inversely proportional to the strength of the magnetic field, while for the latter it is inversely proportional to its square-root.

A large body of literature has already focused on the effects of a steady magnetic field imposed on a shear flow near rigid walls. The most dramatic consequence of the magnetic field is, when it is strong enough, an effective or quasi two-dimensionalisation of the flow \citep{moreau1990magnetohydrodynamics,potherat2000effective} referred to as \textit{Q2D}. This is expected and observed in practice outside boundary layers once the interaction parameter, which characterizes the ratio of Lorentz to inertial forces, becomes large compared to unity.
For weaker magnetic fields turbulent and transitional shear flows typically feature coherent structures such as streamwise streaks, like their non-magnetohydrodynamic counterpart, but their range of existence in terms of $\Rey$ differs. Nevertheless from the point of view of transition to turbulence, they remain subcritical so that again a mismatch between the energy stability threshold $\Rey_E$ and the proper transition values is expected. Moreover as in other shear flows, the underlying non-normality is strong, which results in strong amplification by transient growth mechanisms even without any instability of the base flow.

Most energy stability calculations have been done for very simple flow geometries. The earliest calculations were performed in plane channel geometries for planar Couette and Poiseuille flow \citep{joseph1969stability}.
In the context of MHD flows amenable to the low--$\Rey_m$ approximation, the energy stability of the Hartmann layer has been studied by \cite{lingwood1999stability}.
Idealised geometries such as channel and boundary layer are never found neither in Nature nor even in industrial contexts. We therefore decided to investigate the more realistic rectangular duct geometry, when the applied magnetic field is parallel to one of the sidewalls. This flow has been the subject of several experimental \citep{hartmann1937paper,murgatroyd1953cxlii,moresco2004experimental} and numerical studies \citep{kobayashi2008large,krasnov2010optimal,krasnov2012numerical,krasnov2013patterned,zikanov2014patterned,krasnov2015patterned}. Yet to our knowledge it has never been documented from the point of view of energy stability.

Duct geometries have long been used as research laboratories for the generalisations of linear/nonlinear concepts first developed in channel geometries. In the context of transitional flows, instability threshold \citep{tatsumi1990stability,tagawa2019linear} transient growth \citep{krasnov2010optimal,cassells2019three}, edge states \citep{biau2008transition,brynjell2022edge}, exact coherent states \citep{wedin2009three,uhlmann2010traveling} have been recently documented in square duct geometries. The goal of the present paper is to estimate numerically and report values of $\Rey_E$ for rectangular ducts as functions of both the aspect ratio and the intensity of the magnetic field. Besides this exhaustive parametric study, this study also aims at caracterising physically the coherent structures reported for these parameters, their symmetries and their implication for transition to turbulence at higher Reynolds number.

The paper is structured as follows. The mathematical formulation of the continuous problem is given in \S\ref{sec:problemdefinition}, together with the details about the numerical techniques \mbr{(see also Appendix \ref{sec:appendix_matrices})}.
Results relevant to the channel geometry are given in \S\ref{sec:channel_results}.
Duct results are shown in \S\ref{sec:duct_results} for the non-MHD case and in \S\ref{sec:structures} for the MHD case. Conclusions and outlooks are given in \S\ref{sec:conclusions}.

\section{Problem formulation}
\label{sec:problemdefinition}

Our aim is to model the flow of liquid metal in a periodic duct geometry with four sidewalls.
The flow is subject to a magnetic field imposed in a direction transverse to the flow and parallel to one of the walls.
For simplicity, we focus on the case where the walls are all electrically insulating.
%\yd{\sout{The boundary conditions for the velocity are no slip at each wall, and periodicity along the streamwise direction.}} \yd{YD : see my next comment.}

\subsection{Governing equations}

The flow is governed by the incompressible Navier-Stokes equations for the velocity field, coupled to the Maxwell's equations for the magnetic part. The
quasistatic approximation holds if the magnetic Reynolds number $\Rey_m$ is negligible with respect to unity, which will be assumed throughout the whole paper. In this low-$\Rey_m$ approximation \citep{muller2001magnetofluiddynamics} the induced electric field can be represented as the gradient of the electric potential, determined by  Ohm's law for a moving conductor in combination with Amp\`ere's law, which requires the induced current density to be solenoidal. The original coupled system of equations reads

\begin{align}
  \label{eq:momentum}
  \frac{\partial \bm{u}}{\partial t} + (\bm{u} {\cdot \nabla})\bm{u} &=
  - {\nabla p} + \frac{1}{\Rey} {\nabla}^2 \bm{u}
  + \frac{\Ha^2}{\Rey} \left(\bm{j}\times \bm{e}_B\right),\\
  \label{eq:mass}
  \nabla \cdot \bm{u}&= 0,\\
  \label{eq:Ohm}
  \bm{j}&=-\nabla \phi  + \bm{u} \times \bm{e}_B,\\
  \label{eq:Ampere}
  \nabla \cdot \bm{j} &=0 \leftrightarrow  \nabla^2{\phi} = \nabla \cdot ({ \bm{u} \times  \bm{e}_B}).
\end{align}
The variables $p$ and $\phi$ denote the pressure and electric potential, respectively, whereas ${\bm u}$ denotes the velocity field, $\bm{e}_B$ is the direction of the magnetic field and ${\bm j}$ the electric current density. All quantities are non-dimensionalised using the centerline velocity $U_{c}$ of the laminar flow for velocities, the shorter half-width $H$ of the duct for lengths, the strength $B_0$ of the imposed magnetic field, and the electrical conductivity $\sigma$ of the fluid. This leads to a division by $\rho U_c^2$ for the pressure,
by $U_cB_0H$ for the electric potential and by $\sigma U_cB_0$ for the electric current density. The governing non-dimensional control parameters are the Reynolds number
\begin{equation}
\label{Rey}
 	\Rey \equiv \frac{U_c H}{\nu},
\end{equation}
and the Hartmann number
\begin{equation}
\label{Hart}
\Ha \equiv B_0 H\sqrt{\frac{\sigma}{\rho\nu}},
\end{equation}
where $\rho$ is the fluid density and $\nu$ is its kinematic viscosity.
The walls are electrically insulating, \ie the wall-normal component of the electric current density is zero at each wall. Besides the no-slip condition ${\bm u=0}$ is applied at each wall.

For the energy stability analysis, the flow is first decomposed, according to ${\bm u}={\bm U} + {\bm u}'$, into the base laminar state with parallel velocity field $\bm{U}$ and a perturbation velocity field $\bm{u}'$. Moreover, a similar decomposition leads to the perturbation current density ${\bm j}'$, the perturbation electric potential $\phi'$ and the perturbation pressure $p'$.
The equations for the perturbation fields $\bm{u}'$, $\phi'$ and $p'$ are
\begin{align}
  \label{eq:momentum-pert}
  \frac{\partial \bm{u}'}{\partial t} + (\bm{U} {\cdot \nabla})\bm{u}' +  (\bm{u}' {\cdot \nabla})\bm{U}+(\bm{u}' {\cdot \nabla})\bm{u}'
  &= - {\nabla p}' + \frac{1}{\Rey} {\nabla}^2 \bm{u}'
  + \frac{\Ha^2}{\Rey} \left(\bm{j}'\times \bm{e}_B\right),\\
  \label{eq:mass-pert}
  \nabla \cdot \bm{u}'&= 0,\\
  \label{eq:Ohm-pert}
  \bm{j}'&=-\nabla \phi'  + \bm{u}' \times \bm{e}_B,\\
  \label{eq:Ampere-pert}
  \nabla^2{\phi}' &= \nabla \cdot ({ \bm{u}' \times  \bm{e}_B}).
\end{align}
where the boundary conditions for $ \bm{u}'$ are of Dirichlet type except at the inlet and outlet where periodicity is imposed.
Superscript primes will be dropped from the perturbation quantities throughout the rest of this paper.

\subsection{Duct and channel geometries}

\begin{figure}
  \centering
  \includegraphics[width=0.8\linewidth]{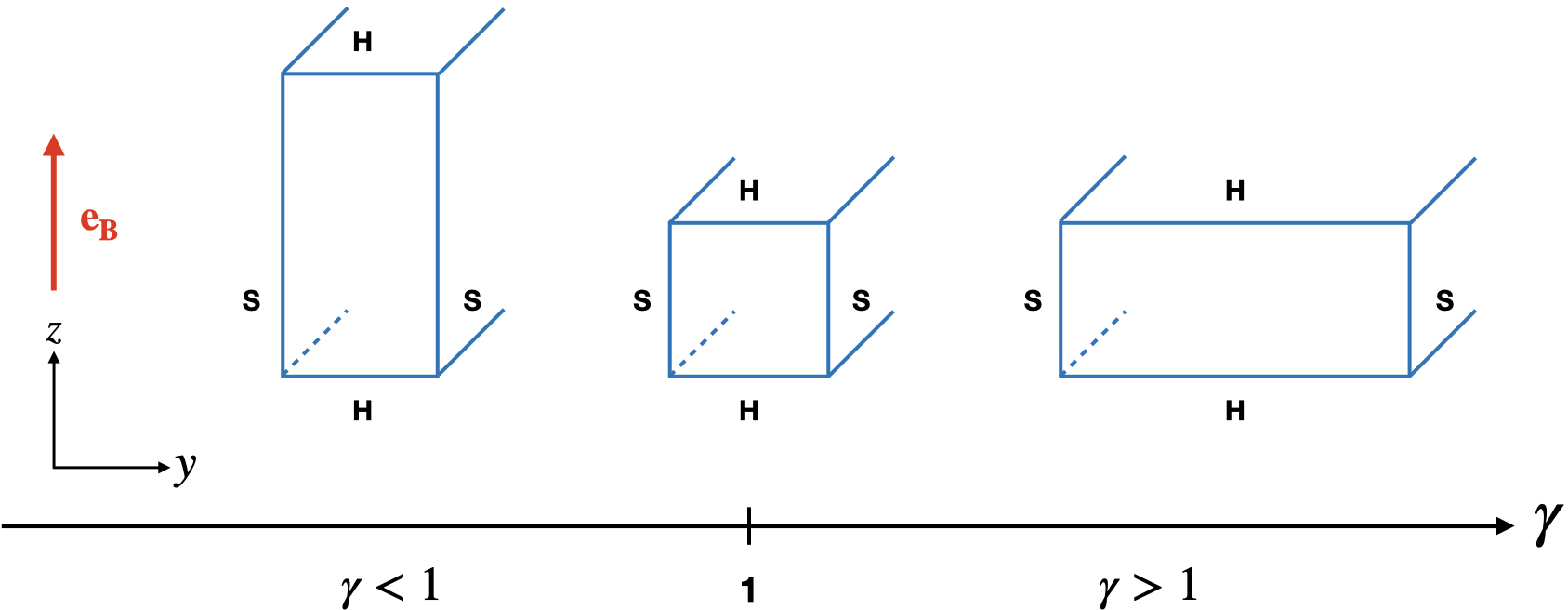}
  \caption{Sketch of the geometry of a cross-section of duct flow, as the aspect ratio $\gamma=L_y/L_z$ is varied from below to above unity ($\gamma=1$ for square duct). The labels $H$ and $S$ stand respectively for the Hartmann and the Shercliff layers in the presence of a magnetic field aligned with the $z$-direction also noted ${\bm e}_B$.  The reference length (used \eg in the definition of the Hartmann number) is always  \tb{one half} of the shorter side: for $\gamma<1$ it is half of the distance between the Shercliff walls and for $\gamma>1$ it is half of the distance between the Hartmann walls. The channel case corresponds to the limit $\gamma \rightarrow 0$.}

  \label{fig:sketch}
\end{figure}

The main geometry under consideration in this study is a duct aligned with the streamwise direction ${\bm x}$. The sides of the cross-section are parallel to the transverse directions
${\bm y}$ and ${\bm z}$.
By convention, the magnetic field is aligned with the ${\bm z}$ direction, $\bm{e}_B=\bm{e}_z$ (this renders the coordinates $y$ and $z$ equivalent in the absence of magnetic field).
The velocity field is considered periodic in the streamwise direction with a period $L_x$. The distance between the sidewalls is noted respectively $\tb{2} L_y$ (resp.~$ \tb{2} L_z$) in the $y$ (resp.~$z$) direction. The reference length $H$, used to build for instance the Reynolds number and the Hartmann number, is always taken to be half the shorter side of the cross-section.
The pedagogic sketch in figure \ref{fig:sketch} explains how the geometry of the cross-section changes from $\gamma<1$ to $\gamma>1$, with $\gamma=1$ referring to a square duct. %The velocity field obeys no slip at all sidewalls, which are considered electrically insulating. The flow moves with a constant flow rate directly proportional to the velocity scale $U_c$.

For the channel geometry, periodicity is assumed both in the streamwise and spanwise direction, which is called here $y$. The reference length becomes the gap between the two walls, while the reference velocity $U_c$ is still the laminar centerline velocity.

\subsection{Base flow}

The base flow is streamwise-independent and only the streamwise velocity component is non-zero. The induced current density is therefore two-dimensional and can be represented by the induced streamwise magnetic field through Amp\`ere's law, $\bm{J}=\nabla\times(B\bm{e}_x)$.
The governing equations in dimensional form are
\begin{align}
  \label{eq:baseflow1}
  \varrho \nu \nabla^2 U +  \frac{B_0}{\mu_0}\frac{\partial B}{\partial z} &=\frac{\partial p}{\partial x}, \\
  \label{eq:baseflow2}
  \lambda_m \nabla^2 B + B_0\, \frac{\partial U}{\partial z}&= 0,
\end{align}
where $\mu_0$ is the magnetic permeability of free space and $\lambda_m =1/(\mu_0\sigma)$ is the magnetic diffusivity. The gradient operators have to be understood as two-dimensional gradients defined with respect to the cross-flow variables $y$ and $z$ only.
By choosing the shorter edge as lengthscale and appropriate units for  $U$ and $B$, one can make the prefactors of the terms multiplying the $z$-derivatives on the  left hand sides equal and the pressure gradient equal to unity. The non-dimensional equations for the base flow then read
\begin{align}
  \label{eq:baseflow1_nd}
  \nabla^2 U +  \Ha\, \frac{\partial B}{\partial z} &=1, \\
  \label{eq:baseflow2_nd}
  \nabla^2 B + \Ha\, \frac{\partial U}{\partial z} &=0.
\end{align}
These equations can be decoupled by adding and subtracting them. One obtains the two equations
\begin{equation}
  \label{eq:elsaesser}
  \left(\nabla^2\pm \Ha\, \frac{\partial }{\partial z}\right) \left(U\pm B\right)=1,
\end{equation}
for the Shercliff variables $U\pm B$ with homogeneous Dirichlet conditions.

\begin{figure}
  \centering
  \begin{subfigure}[b]{0.49\textwidth}
    \footnotesize (a) \hspace{-1mm}
    \includegraphics[width=\linewidth,clip=]{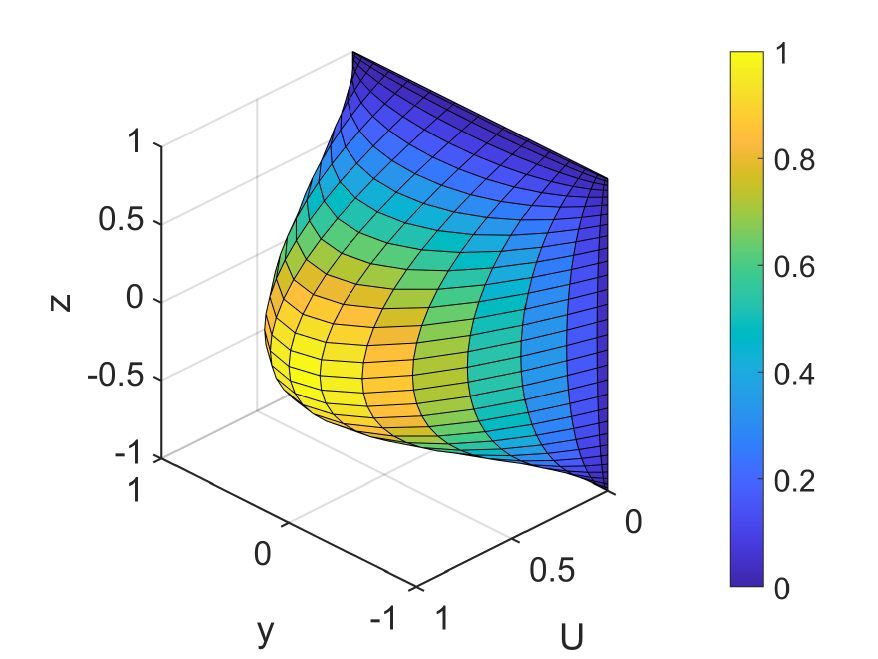}
    \phantomsubcaption
    \label{fig:bf_ha0_beta1}
  \end{subfigure}
  \hfill
  \begin{subfigure}[b]{0.49\textwidth}
    \footnotesize (b) \hspace{-1mm}
    \includegraphics[width=\linewidth,clip=]{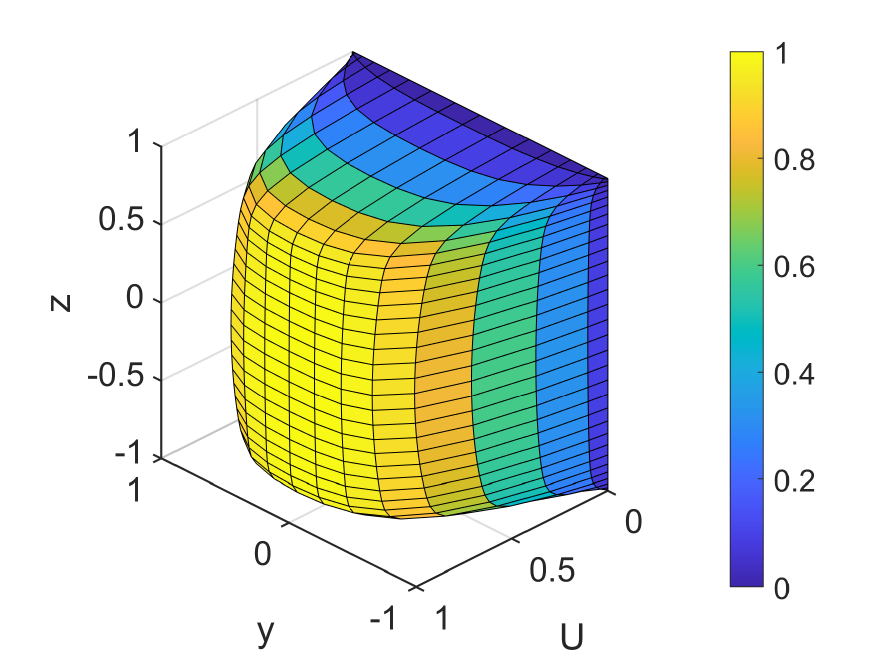}
    \phantomsubcaption
    \label{fig:bf_ha20_beta1}
  \end{subfigure}\\
  \begin{subfigure}[b]{0.49\textwidth}
    \footnotesize (c) \hspace{-1mm}
    \includegraphics[width=\linewidth,clip=]{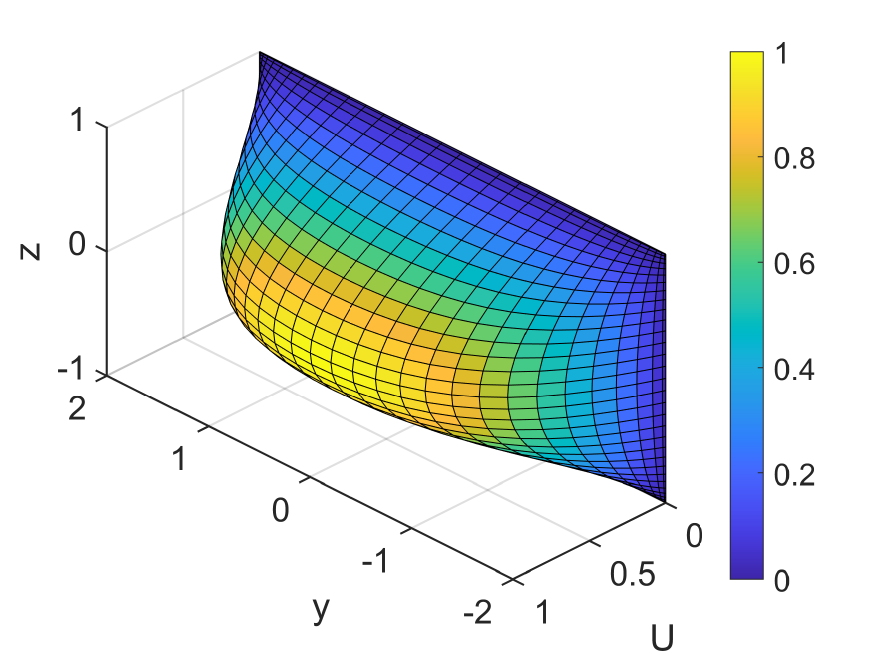}
    \phantomsubcaption
    \label{fig:bf_ha0_beta2}
  \end{subfigure}
  \hfill
  \begin{subfigure}[b]{0.49\textwidth}
    \footnotesize (d) \hspace{-1mm}
    \includegraphics[width=\linewidth,clip=]{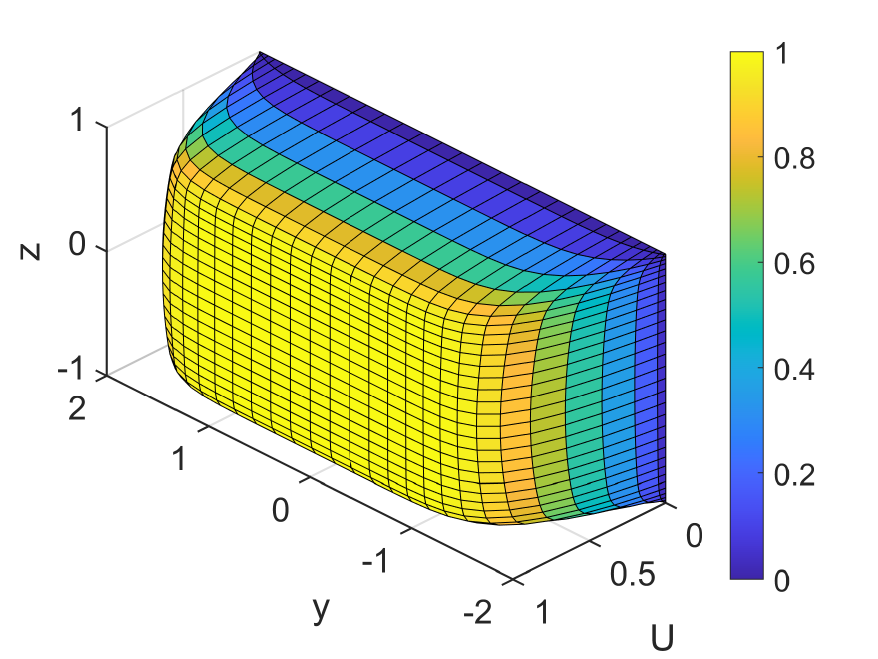}
    \phantomsubcaption
    \label{fig:bf_ha20_beta2}
  \end{subfigure}
  \caption{Base flow for $\Ha$=0 (left) and $\Ha$=20 (right) for an aspect ratio $\gamma=$1 (top) and $\gamma=2$ (bottom). Isosurface of the streamwise velocity $U(y,z)$.}
  \label{fig:figure2}
\end{figure}

\begin{figure}
  \centering
  \begin{subfigure}[b]{0.32\textwidth}
    \footnotesize (a) \hspace{-1mm}
    \includegraphics[width=0.9\linewidth,clip=]{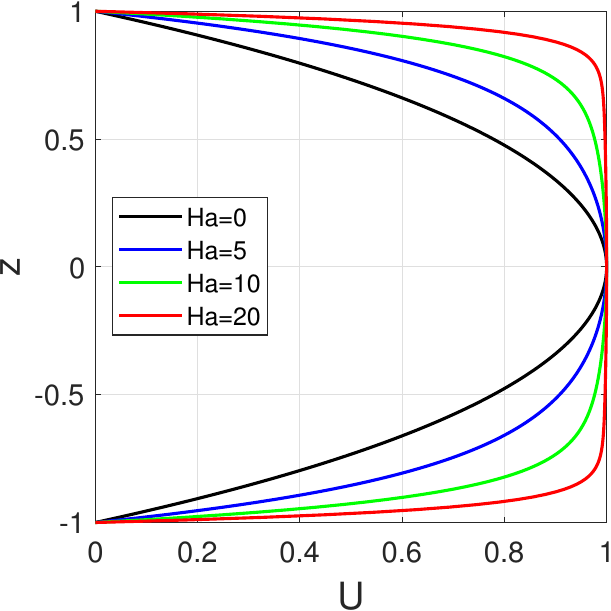}
    \phantomsubcaption
    \label{fig:bf_u_z_profile}
  \end{subfigure}
  \hfill
  \begin{subfigure}[b]{0.32\textwidth}
    \footnotesize (b) \hspace{-1mm}
    \includegraphics[width=0.9\linewidth,clip=]{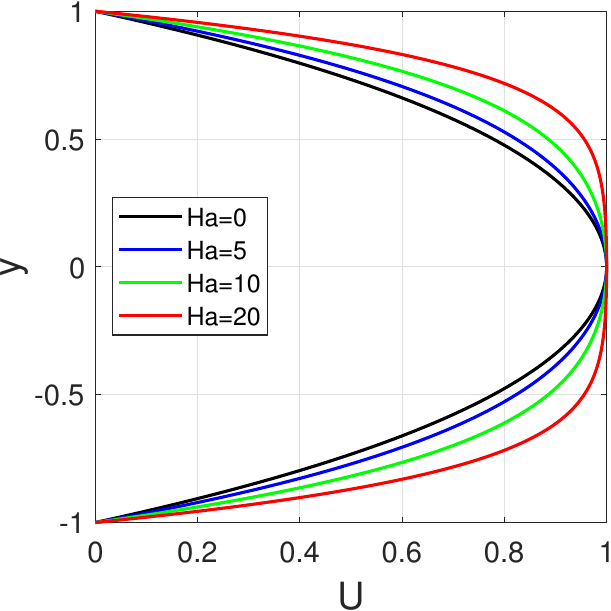}
    \phantomsubcaption
    \label{fig:bf_u_y_profile}
  \end{subfigure}
  \hfill
  \begin{subfigure}[b]{0.32\textwidth}
    \footnotesize (c) \hspace{-1mm}
    \includegraphics[width=0.9\linewidth,clip=]{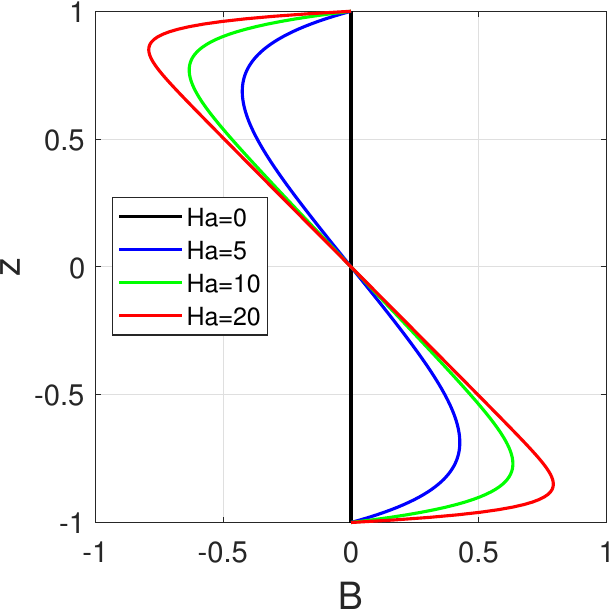}
    \phantomsubcaption
    \label{fig:bf_b_z_profile}
  \end{subfigure}
  \caption{Profiles of the base flow in the midplanes for different Hartmann numbers and $\gamma=1$. (\subref{fig:bf_u_z_profile}) $z$-profile of the streamwise velocity, (\subref{fig:bf_u_y_profile}) $y$-profile of the streamwise velocity, (\subref{fig:bf_b_z_profile}) $z$-profile of the streamwise magnetic flux density.}
  \label{fig:baseflow_1d_profiles}
\end{figure}

An analytical solution to eq.~\eqref{eq:elsaesser} in the form of a Fourier series was originally derived by \cite{shercliff_1953} and later elaborated upon in \citep{muller2001magnetofluiddynamics}.
However, in this work \eqref{eq:elsaesser} is discretised as described in \S\ref{ch:spatial_discretisation} and solved directly.
Upon resolution, the desired base velocity distribution is obtained from the sum of the appropriately scaled Shercliff variables.
This solution is shown in figure \ref{fig:figure2} \mbr{and \ref{fig:baseflow_1d_profiles}} for different duct aspect ratio $\gamma$ and Hartmann numbers. The Hartmann and Shercliff layers on the walls $z=\pm 1$ and $y=\pm \gamma$ are clearly apparent by comparison between the cases $\Ha=0$ and $\Ha=20$.

\subsection{Energy stability as a minimisation problem}
Energy stability analysis  is concerned with the behavior of  the total perturbation kinetic energy, defined as
\begin{equation*}
  E = \int_V  \frac{u_i^2}{2} \, dV,
\end{equation*}
using tensor notation.
The evolution equation for $E$ is obtained by multiplying the momentum equation \eqref{eq:momentum-pert} by $\bm{u}$ and integrating over the volume of the duct.
Using  integration by parts, this leads to
\begin{equation}
  \frac{\partial E}{\partial t} = -\int_V u_i u_l \frac{\partial U_i}{\partial x_l} \, dV- \frac{1}{\Rey} \int_V  \frac{\partial u_i}{\partial x_l} \frac{\partial u_i}{\partial x_l}\, dV - \frac{\Ha^2}{\Rey}\int_V  j_ij_i \, dV.
\end{equation}
This equation is equivalent to the Reynolds-Orr equation \citep{reddy1993energy} in wall-bounded shear flows, save for the additional contribution of the Lorentz force. The slowest possible temporal decay of $E$ occurs for the perturbation that provides the minimum of the functional \citep{doering1995applied}
\begin{equation}
  \frac{1}{E} \left(\int_V u_i u_l \frac{\partial U_i}{\partial x_l} \, dV+ \frac{1}{\Rey} \int_V  \frac{\partial u_i}{\partial x_l} \frac{\partial u_i}{\partial x_l}\, dV + \frac{\Ha^2}{\Rey}\int_V  j_ij_i \, dV\right).
\end{equation}
This functional is subject to the constraint \eqref{eq:mass-pert}, and the current density is represented by \eqref{eq:Ohm-pert} and \eqref{eq:Ampere-pert}. We use Lagrange multipliers $q$ and $\lambda$  to add the mass conservation and energy normalization constraints to the functional, \ie we seek the extrema of the scalar functional $F$, defined by
\begin{align}
  F &= \int_V u_i u_l \frac{\partial U_i}{\partial x_l} \, dV+ \frac{1}{\Rey} \int_V  \frac{\partial u_i}{\partial x_l} \frac{\partial u_i}{\partial x_l}\, dV + \frac{\Ha^2}{\Rey}\int_V  j_ij_i \, dV\nonumber\\
  &-  \int_V q\, \nabla \cdot \bm{u}\, dV -\lambda \left(E -1\right),
\end{align}
where the minimisation is carried over all admissible divergence-free velocity fields satisfying the boundary conditions.
The current density $\bm{j}$ depends directly on ${\bm u}$ and is given by equation \eqref{eq:Ohm-pert} with $\phi$ satisfying equation \eqref{eq:Ampere-pert}. One stationarity condition is obtained via variation of the velocity field, \ie from
\begin{equation*}
  0=\frac{\delta F}{\delta \bm{u}}=\frac{d}{d\varepsilon}\left. F[\bm{u}+\varepsilon \delta \bm{u}]\right|_{\varepsilon=0}.
  \label{eq:extremumcondition}
\end{equation*}
It leads to the Euler-Lagrange equation
\begin{equation}
  \label{eq:energystability1}
  \lambda  \bm{u} =-\nabla q + 2 \bm{\hat{S}}\cdot \bm{u} -   \frac{2}{\Rey} \nabla^2 \bm{u} -  \frac{2 \Ha^2}{\Rey} \bm{j}\times \bm{e}_B,
\end{equation}
where $ \bm{\hat{S}}$ is the symmetric part of the velocity gradient of the base flow and the current density is defined through equations \eqref{eq:Ohm-pert}-\eqref{eq:Ampere-pert}. Variation of $F$ with respect to $q$ gives the constraint \eqref{eq:mass-pert}. The multiplier $\lambda$ is the growth rate of the perturbation satisfying equations \eqref{eq:energystability1}, \eqref{eq:mass-pert}, \eqref{eq:Ohm-pert} and \eqref{eq:Ampere-pert} for given values $\Rey$ and $\Ha$.
We are interested in the lowest value of $\Rey$ where non-decaying solutions exist, \ie $\lambda=0$. The minimizing velocity field is such that equation \eqref{eq:energystability1} reduces to the following eigenvalue problem for $\Rey$ \citep{doering1995applied}:
\begin{equation}
  \label{eq:energystability2}
  \Rey\, \bm{\hat{S}}\cdot \bm{u} =  -\frac{\nabla q}{2} +  \nabla^2 \bm{u} +  \Ha^2 \bm{j}\times \bm{e}_B.
\end{equation}
%where $p$ is the pressure field $q$ \yd{rescaled by a factor of two}.
The lowest eigenvalue $\Rey$ defines the energy stability Reynolds number $\Rey_E$. The corresponding eigenvector represents a flow field whose kinetic energy does, for $\Rey=\Rey_E$, neither experiences initial growth nor initial decay.
The spectral problem \eqref{eq:energystability2} admits other eigenvalues beyond the lowest one.
They correspond to larger values of $\Rey$ for which the problem admits neutral modes, \ie non-monotonically decaying energy variations. By convention, each eigenvector indexed by $i=1,2,...$ corresponds to the neutral flow field expressed at the value of $\Rey=\Rey^{(i=1,2,...)}$ at which it is neutral.

\subsection{Detailed formulation}

The incompressiblity condition  leads to difficulties for the numerical solution of the energy stability equations. We therefore adopt the approach used by \cite{priede2010linear} and represent the velocity field by a vector stream function $\bm{\psi}$, i.e.
\begin{equation}
  \label{eq:vectorstreamfunction}
  \bm{u}=\nabla \times \bm{\psi}.
\end{equation}
By that, $\nabla\cdot\bm{u}=0$ is always satisfied. The vector streamfunction is defined only up to an additive gradient field. In order to fix this gradient field, we impose the gauge condition
\begin{equation}
  \label{eq:coulomb}
  \nabla \cdot \bm{\psi}=0,
\end{equation}
whereby $\bm{\psi}$ is defined up to the gradient of a harmonic function. The condition \eqref{eq:coulomb} also simplifies the relation between $\bm{\psi}$ and the vorticity field  to $\bm{\omega}=-\nabla^2\bm{\psi}$.

By taking the curl of equation \eqref{eq:energystability2}, we obtain equations for $\omega_y$ and $\omega_z$ and \yd{eliminate the field $q$}. They read
\begin{align}
  \label{eq:estab-vectorsf1}
  \nabla^2 \omega_y -\Ha^2 \frac{\partial }{\partial z}\left(\frac{\partial \phi}{\partial y}+u_x\right) &= \Rey\, \bm{e}_y\cdot\nabla\times \, \bm{\hat{S}}\cdot \bm{u}, \\
  \label{eq:estab-vectorsf2}
  \nabla^2 \omega_z -\Ha^2 \frac{\partial^2 \phi}{\partial z^2} &= \Rey\, \bm{e}_z\cdot\nabla\times \, \bm{\hat{S}}\cdot \bm{u},\\
  \label{eq:estab-vectorsf3}
  \nabla^2 \psi_y+ \omega_y   &= 0,\\
  \label{eq:estab-vectorsf4}
  \nabla^2 \psi_z+ \omega_z   &= 0,\\
  \label{eq:estab-vectorsf5}
  \nabla^2 \phi- \omega_z   &= 0.
\end{align}

\subsubsection{Streamwise-dependent perturbations}

Since the streamwise direction is homogeneous, the eigenfunctions of the energy stability eigenvalue problem are Fourier modes with streamwise wavenumber $\alpha$. We therefore write
\begin{equation}
\label{eq:alpha}
 \left\{\bm{\psi},\bm{\omega},\phi\right\}(x,y,z)= \left\{\bm{\hat{\psi}}(y,z),\bm{\hat{\omega}}(y,z),\hat{\phi}(y,z)\right\}e^{i\alpha x}.
\end{equation}
Equations \eqref{eq:estab-vectorsf3}-\eqref{eq:estab-vectorsf5} turn into three two-dimensional Helmholtz equations for  each Fourier mode of the components $\psi_y$, $\psi_z$ and the electric potential $\phi$  with the vorticity components as right hand sides. Each of them is supplemented with a homogeneous boundary condition. Upon discretization, these equations become linear invertible mappings between the discrete representations of $\psi_y$, $\psi_z$ and $\phi$ and the discrete representations of $\omega_y$, $\omega_z$ augmented by a set of zero boundary  data. The actual eigenvalue problem consists of
equations \eqref{eq:estab-vectorsf1}-\eqref{eq:estab-vectorsf2} with $\omega_y$ and $\omega_z$ as independent variables.
With this representation, the streamwise components $\psi_x$ and $\omega_x$ are directly obtained from the other two components $\psi_y, \psi_z$ and $\omega_y,\omega_z$ via equation \eqref{eq:coulomb} and $\nabla\cdot\bm{\omega}=0$.

The boundary conditions for the Fourier modes of $\psi_y$, $\psi_z$ or  $\omega_y$, $\omega_z$ have to be formulated such that the  no-slip condition is satisfied.
Following \cite{priede2010linear}, we first impose that the tangential vector streamfunction component  ${\psi}_t$ in the $(y,z)$ plane vanishes on each wall. This is admissible since it is equivalent to a Dirichlet condition for the arbitrary harmonic function (whose gradient can be added to  $\bm{\psi}$).  The other conditions are $u_x=0$ and $u_n=0$, where subscript $n$ denotes the normal component.
We note that $u_n=0$ is equivalent to $\partial {\psi}_n/\partial n=0$ when ${\psi}_t=0$ and that  $u_x=0$ implies $\partial {\psi}_z/\partial y-\partial {\psi}_y/\partial z=0$. The third condition  $u_t=0$ in the $(y,z)$ plane is equivalent to $\omega_n=0$. For equations \eqref{eq:estab-vectorsf3} for $\psi_y$ and \eqref{eq:estab-vectorsf4} for $\psi_z$, one of the conditions $\psi_t=0$ and $\partial {\psi}_n/\partial n=0$ is selected on each segment of the boundary. Equation \eqref{eq:estab-vectorsf5} for $\phi$ requires homogeneous Neumann conditions.
Equations \eqref{eq:estab-vectorsf1}-\eqref{eq:estab-vectorsf2} are complemented with the conditions $\partial {\psi}_z/\partial y-\partial {\psi}_y/\partial z=0$ or $\omega_n=0$.

\subsubsection{Streamwise-independent perturbations}

The case of zero streamwise wavenumber  must be treated separately since the representation of the streamwise velocity $u_x$ by the other components fails when $\alpha=0$. One can then use the classical scalar stream function representation
\begin{subequations}
  \label{eq:streamfunction}
  \begin{gather}
    u_y=\frac{\partial \psi_x}{\partial z}, \qquad
    u_z=-\frac{\partial \psi_x}{\partial y}.
    \tag{\theequation a,b}
  \end{gather}
\end{subequations}
for the in-plane components of the velocity. Likewise, the in-plane components of the electric current density are
\begin{subequations}
  \label{eq:bx}
  \begin{gather}
    j_y=\frac{\partial \chi}{\partial z}, \qquad
    j_z=-\frac{\partial \chi}{\partial y},
    \tag{\theequation a,b}
  \end{gather}
\end{subequations}
where the stream function $\chi$ for the current density is proportional to  the streamwise component of the induced magnetic field. The streamwise current density is $j_x=u_y$. It stems from Ohm's law with the assumption that a mean streamwise current is excluded. Using Ohm's law one can also show that
\begin{equation}
  \label{eq:bx2}
  \nabla^2 \chi=-\frac{\partial u_x}{\partial z}.
\end{equation}
Equations for $\psi_x$, $\omega_x$, $u_x$ and $\chi$ are obtained from the streamwise component of the equation \eqref{eq:energystability2} and the streamwise component of its curl. These equations are
\begin{align}
  \label{eq:estab-scalarsf1}
  \nabla^2 u_x + \Ha^2 \frac{\partial \chi}{\partial z} &= \frac{\Rey}{2}\, \left(\frac{\partial U}{\partial y}\frac{\partial \psi_x}{\partial z}-
  \frac{\partial U}{\partial z}\frac{\partial \psi_x}{\partial y}\right), \\
  \label{eq:estab-scalarsf2}
  \nabla^2 \omega_x - \Ha^2 \frac{\partial^2 \psi_x}{\partial z^2} &= \frac{\Rey}{2}\, \left(\frac{\partial U}{\partial y}\frac{\partial u_x}{\partial z}-
  \frac{\partial U}{\partial z}\frac{\partial u_x}{\partial y}\right), \\
  \label{eq:estab-scalarsf3}
  \nabla^2 \psi_x -\omega_x &= 0,\\
  \label{eq:estab-scalarsf4}
  \nabla^2 \chi + \frac{\partial u_x}{\partial z} &= 0.
\end{align}
The boundary conditions for equation \eqref{eq:estab-scalarsf3} and \eqref{eq:estab-scalarsf4} are $\psi_x=0$ and $\chi=0$. For the other two equations the no-slip conditions $u_x=0$ and $u_t=0$ are required. The latter is equivalent to  $\partial {\psi}_x/\partial n=0$.
As discussed earlier for the case $\alpha>0$, the quantities $\chi$ and $\psi_x$ are obtained via one-to-one maps  from $\omega_x$ and $u_x$. The actual eigenvalue problem consists of equations \eqref{eq:estab-scalarsf1} and \eqref{eq:estab-scalarsf2}.

\subsection{Spatial discretisation for the duct and channel geometry}
\label{ch:spatial_discretisation}
We use a spectral collocation method based on the Chebyshev polynomials $T_0,T_1,...$ defined over $[-1,1]$ by $T_n(x)=\cos(n\arccos x), n \ge 0$ \citep{canuto2007spectral}.

For the duct flow, both cross-stream vorticity components  are expanded as a finite double sum, i.e.
\begin{equation}
\label{eq:ansatz}
  f(y,z)=\sum_{k=0}^{N_y}\sum_{l=0}^{N_z} \hat{f}_{k,l}\, T_k(y/L_y)\, T_l(z/L_z),
\end{equation}
where $f$ denotes either $\omega_y$ or $\omega_z$.
Equations \eqref{eq:estab-vectorsf1}-\eqref{eq:estab-vectorsf2} and the corresponding boundary conditions for $\omega_y$, $\omega_z$ are enforced pointwise at the $(N_y+1)(N_z+1)$ Gauss-Lobatto collocation points $y_k$ and $z_l$ defined by
\begin{equation}
\label{eq:glc}
y_k=L_y\cos\left(k\pi/N_y\right),\quad
z_l=L_z\cos\left(l\pi/N_z\right).
\end{equation}
Each collocation point provides one scalar equation for the expansion  coefficients of $\omega_y$ and $\omega_z$ \citep{canuto2007spectral}.
The corners of the rectangular domain may require special consideration when derivatives are specified along the boundaries. In these cases it may be appropriate to impose the differential equation itself at a corner.

As a result, one obtains a generalized linear eigenvalue problem of the type
\begin{equation}
\label{eq:generalizedevp}
\bm{\mathsf{A}} \bm{Y} =\Rey\, \bm{\mathsf{B}} \bm{Y},
\end{equation}
where the vector $\bm{Y}$ contains the unknown expansion coefficients of $\omega_y$ and $\omega_z$.
The streamfunction components and the electric potential in equations \eqref{eq:estab-vectorsf1}-\eqref{eq:estab-vectorsf2} are represented as linear functions of $\omega_y$ or $\omega_z$ since they are given by equations \eqref{eq:estab-vectorsf3}-\eqref{eq:estab-vectorsf5}.
The discrete representation of these quantities is also obtained through spectral collocation. However, this requires more collocation points because $\psi_y$, $\psi_z$ and $\phi$ do not only depend on the inhomogeneity but also on the boundary data.
We therefore use $(N_y+3)(N_z+3)$ expansion coefficients for $\psi_y$, $\psi_z$ and $\phi$ in the ansatz \eqref{eq:ansatz} and in eq. \eqref{eq:glc}. By that, we obtain  an invertible linear system  between the expansion coefficients of either $\omega_y$ or $\omega_z$ augmented by the zero boundary data and the expansion coefficients of $\psi_y$, $\psi_z$ or $\phi$. These three inverse matrices are computed and stored before the matrices of problem \eqref{eq:generalizedevp} are assembled. The computation of these matrices as well as of matrices $\bm{\mathsf{A}}$,  $\bm{\mathsf{B}}$ is described in the Appendix.

Problem \eqref{eq:generalizedevp} was solved with MATLAB's \texttt{eig} routine to find all eigenvalues and eigenvectors. The routine also works with a matrix $\bm{\mathsf{B}}$ whose rank is smaller than the rank of $\bm{\mathsf{A}}$ (as it is the case for \eqref{eq:generalizedevp}). It associates the spurious solutions that stem from equations not containing the eigenvalue $\Rey$ with infinite eigenvalues.
The numerical approach for the special case $\alpha=0$ is analogous with $\omega_x $ and $u_x$ taking the role of $\omega_y$ and $\omega_z$ as primary unknowns.

In contrast to the duct geometry, the base flow in the infinitely wide channel depends only on the $z$-coordinate. Owing to this homogeneity, the solution to the eigenvalue problem can be represented by Fourier modes with respect to $x$ and $y$ with arbitrary wavenumbers $\alpha$ and $\beta$. The ansatz for a velocity or vorticity component then becomes
\begin{equation}
  \label{eq:ansatz2}
  f(x,y,z)=\sum_{k=0}^{N_z} \tilde{f}_{k}\, T_k(z/L_z)\, e^{i\left( \alpha x +\beta y\right)}.
\end{equation}
The velocity field is represented through the vertical velocity and vorticity components. Equations for these quantities are obtained by taking the vertical components of the curl and the double curl of the stability eigenvalue problem \eqref{eq:energystability2}. These equations are complemented by the equation \eqref{eq:Ampere-pert} for the electric potential. The number of unknowns corresponds to the expansion coefficients of vertical velocity, vorticity and potential, \ie approximately $3 N_z$. The discretized form is obtained by enforcing the equations pointwise at collocation points $z_k$, and the resulting generalized linear eigenvalue problem, also of the form \eqref{eq:generalizedevp}, is solved with MATLAB's \texttt{eig} routine.
The energy stability of the Q2D model was formulated in a similar way with the spanwise velocity component as the sole dependent variable.

\subsection{Code verification and numerical resolution}
\mbr{The code has previously been used in the context of magnetoconvection \citep{bhattacharya2024wall}. Its accuracy for the duct geometry was verified with linear stability results from the hydrodynamic literature.}
%The code for the duct geometry was verified with linear stability results from the hydrodynamic literature.
For the streamwise-independent perturbations we computed the eigenvalues of the Stokes operator for $\gamma=1$ and compared them with \cite{leriche2004stokes}. The 10 leading eigenvalues in table 2 of \cite{leriche2004stokes} were reproduced to at least 8 significant digits with a resolution of $N_y=N_z=25$.
For the perturbations with $\alpha>0$ we took a case from \cite{priede2010linear} (their table 2, left column) with a simplified base flow $(1-y^2)(1-z^2)$ in a square duct.  For $\Rey=10^4$, $\alpha=1$ we reproduced the complex relative phase velocity of the leading eigenmode to six significant digits with a resolution of $N_y=N_z=60$ modes.

A direct comparison for energy stability was only possible without magnetic field for $\gamma=1$ (see \S\ref{sec:energy_stability_ha0}).  The additional electromagnetic terms in the equations could not be checked directly. However, the MHD channel results should provide appropriate limits to the duct results for either large $\Ha$ or small/large $\gamma$. This will also become apparent in the following sections.

\begin{table}
  \begin{center}
    \begin{tabular*}{.5\textwidth}{@{\extracolsep{\fill}}cccccc@{}}
    %\begin{tabular}{cccccc}
      {$\Ha$} & {$\gamma$} & {$N_1$} & {$N_2$} & $\Rey_E$ & { $\Delta \Rey_E/\Rey_E$} \\[3pt]
      $0$ & $1$ & $25$ &  $30$ & $77.828$ & $5\times 10^{-8}$\\
      $10$ & $1$ & $30$ &  $35$ & $174.92$ & $2\times 10^{-7}$\\
      $20$ & $1$ & $35$ &  $40$ & $300.59$& $1\times 10^{-8}$\\
      $40$ & $1$ & $45$ &  $53$ & $546.84$ & $6\times 10^{-8}$\\
      $60$ & $1.2$ & $60$ &  $70$ & $792.80$ & $4\times 10^{-8}$\\
    \end{tabular*}
  \end{center}
  \caption{Resolution tests for MHD duct flow with $\alpha=2$. Energy stability eigenvalues $\Rey_E$ were computed with maximum orders $N_1$ and $N_2$ of Chebyshev polynomials  in both $y$ and $z$ resulting in a difference  $\Delta \Rey_E$.  }
  \label{table:accuracy}
\end{table}

The numerical resolution for the duct flow has to be increased with $\Ha$ in order to resolve the electromagnetic boundary layers. The requirements were systematically tested for $\gamma\approx 1$ and different $\Ha$ by comparing two different resolutions. Table \ref{table:accuracy} indicates that the results are sufficiently accurate for the lower order $N_1$ of Chebyshev polynomials. However, the accuracy also depends on $\alpha$. It becomes poorer as $\alpha$ is decreased. This can be expected since $\alpha\to 0$ is a  singular limit for the formulation based on $\omega_y$ and $\omega_z$.

When the  aspect ratio $\gamma$ is not close to unity, the number of polynomials must be increased along the longer edge of the duct to maintain adequate resolution. We decided to keep the maximum spacing of  the collocation points constant on the longer edge. Since this spacing  scales as $1/N$ (where $N$ is the polynomial order), the appropriate choice is to multiply $N_y$ by $\gamma$ or to divide $N_z$ by $\gamma$ (for $\gamma>1$ and $\gamma<1$, respectively). This is done relative to the reference case $\gamma=1$. Depending on the lowest $\alpha$ of interest, the base resolution may have to be increased to ensure valid results. This can also be detected from the magnitude of the imaginary part of $\Rey_E$, which should ideally be  zero. Eigenvalues with significant imaginary parts are discarded in the computations.

The numerical resolution is mainly limited by the computing time, which approximately scales with the third power of the number of unknowns. For $N_y=N_z=60$, the assembly of the matrices and eigenvalue computation  took about 2.5 hours for fixed $\alpha$ and $\gamma$ on an  Intel Xeon E5 processor.

\section{Channel geometry}
\label{sec:channel_results}

\subsection{Energy stability for $\Ha=0$}

We begin by considering the purely hydrodynamic channel case with only two parallel walls, when $\Ha=0$.
This configuration is one of the earliest cases treated in the literature. For a recent comparative review we refer for instance to \cite{falsaperla2019nonlinear}. Orr has initially sought neutral modes under the hypothesis that their spanwise wavenumber $\beta$ is zero and found analytically a value of $\Rey_E\approx 87$ \citep{orr1907stability}. A more accurate value is $\Rey_E= 87.6$ at $\alpha= 2.09$ \citep{falsaperla2019nonlinear}.
Later \cite{joseph1969stability,busse1969bounds}, by seeking neutral perturbations with zero streamwise wavenumber $\alpha$, reported a lower value of $\Rey_E= 49.6$ at $\beta = 2.04$. In the present computation, both $\alpha$ and $\beta$
can be freely varied. The two values of $\Rey_E$ put forward by Orr and by \cite{busse1969bounds} are confirmed in figure \ref{fig:channelHa0} by focusing on the axes $\alpha=0$ or $\beta=0$.
Whereas the value for $\alpha=0$ corresponds to a local minimum of $\Rey_E$ in the $(\alpha,\beta)$ plane, the minimiser for $\beta=0$ appears as a saddle in
the unfolded $(\alpha,\beta)$ plane.

\subsection{Influence of increasing $\Ha$}

\begin{figure}
  \centering
  \begin{subfigure}[b]{0.49\textwidth}
    \footnotesize (a) \hspace{-1mm}
    \includegraphics[width=0.9\linewidth]{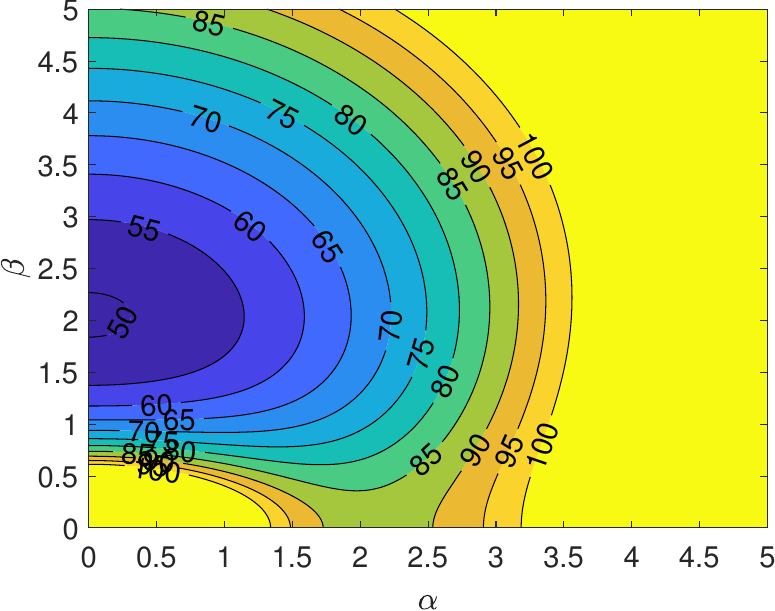}
    \phantomsubcaption
    \label{fig:channelHa0_a}
  \end{subfigure}
  \hfill
  \begin{subfigure}[b]{0.49\textwidth}
    \footnotesize (b) \hspace{-1mm}
    \includegraphics[width=0.9\linewidth]{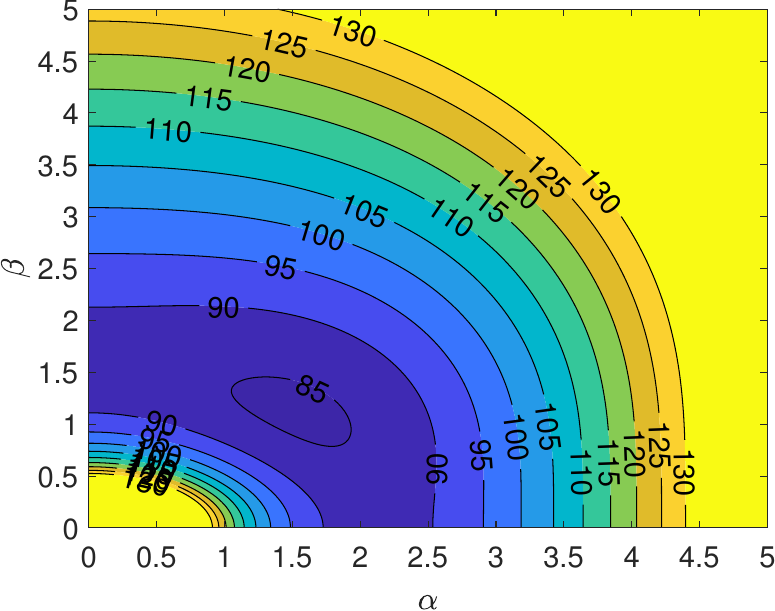}
    \phantomsubcaption
    \label{fig:channelHa0_b}
  \end{subfigure}\\
  \begin{subfigure}[b]{0.49\textwidth}
    \footnotesize (c) \hspace{-1mm}
    \includegraphics[width=0.9\linewidth]{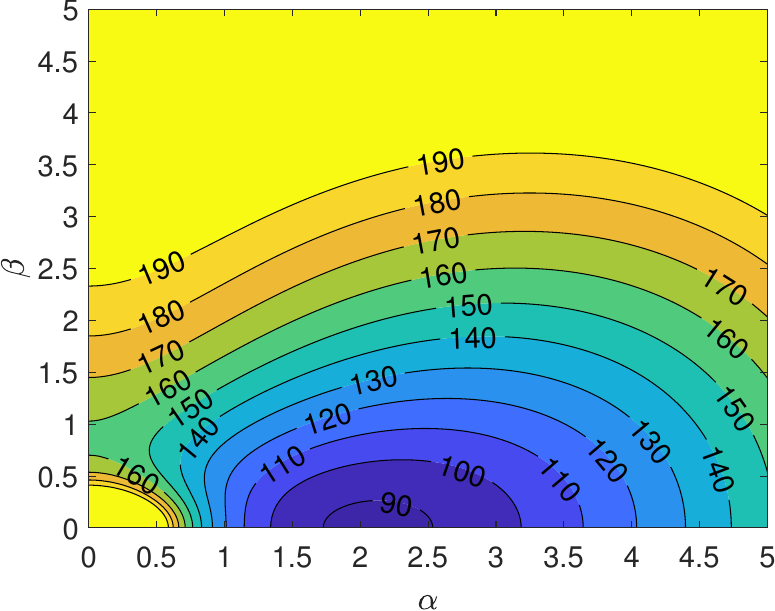}
    \phantomsubcaption
    \label{fig:channelHa0_c}
  \end{subfigure}
  \hfill
  \begin{subfigure}[b]{0.49\textwidth}
    \footnotesize (d) \hspace{-1mm}
    \includegraphics[width=0.9\linewidth]{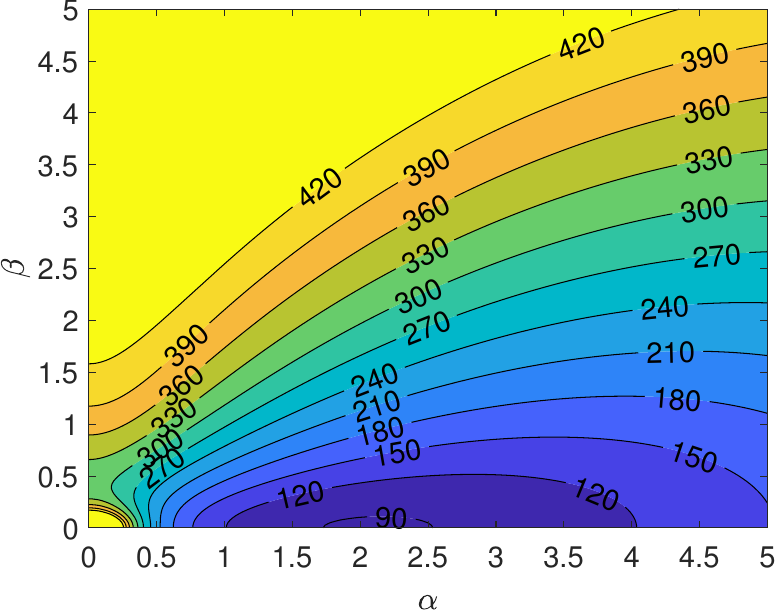}
    \phantomsubcaption
    \label{fig:channelHa0_d}
  \end{subfigure}
  \caption{Cartography of $\Rey_E$ in the $(\alpha,\beta)$ plane, where $\alpha$ and $\beta$ are respectively the streamwise and spanwise wavenumber. Channel geometry, from (\subref{fig:channelHa0_a}) to (\subref{fig:channelHa0_d}) $\Ha=0,5,10,20$.}
  \label{fig:channelHa0}
\end{figure}

The local minima of the $\Rey_E$ in the $(\alpha,\beta)$ plane evolve as $\Ha$ departs from zero. Maps of $\Rey_E$ can be seen in figure \ref{fig:channelHa0} for $\Ha=5,10$ and $20$. For $\Ha \ge 10$ the global minimiser for $\Rey_E$ corresponds to a mode with $\beta=0$ in strong contrast with the case $\Ha=0$.  This minimiser is actually independent of $\Ha$ since there is no Lorentz force for $\beta=0$. It represents the solution found by Orr.
For the intermediate value $\Ha=5$ the minimser is neither along the axis $\alpha=0$ nor along the axis $\beta=0$. Instead it corresponds to an oblique wave vector with both $\alpha$ and $\beta$ non-zero.

\section{Rectangular Duct geometry}
\label{sec:duct_results}

We move now to the rectangular duct case with four walls and a transverse magnetic field parallel to one of the walls. The minimisation problem leading to the value of $\Rey_E$ is governed by two main parameters, notably the aspect ratio $\gamma=L_y/L_z$, and the Hartmann number $\Ha$ based on the shorter edge.

\begin{figure}
  \centering
  \begin{subfigure}[b]{0.49\textwidth}
    \footnotesize (a) \hspace{-1mm}
    \includegraphics[width=0.9\linewidth]{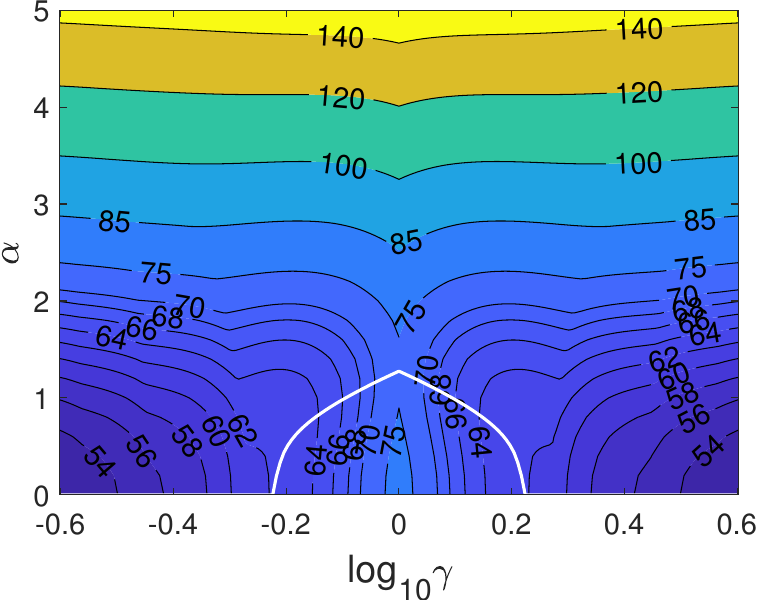}
    \phantomsubcaption
    \label{fig:ductHa0_a}
  \end{subfigure}
  \hfill
  \begin{subfigure}[b]{0.49\textwidth}
    \footnotesize (b) \hspace{-1mm}
    \includegraphics[width=0.9\linewidth]{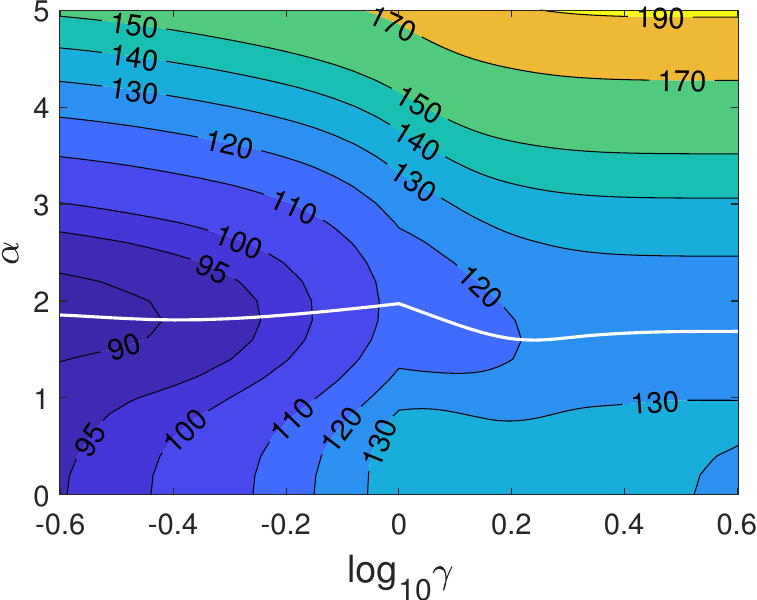}
    \phantomsubcaption
    \label{fig:ductHa0_b}
  \end{subfigure}\\
  \begin{subfigure}[b]{0.49\textwidth}
    \footnotesize (c) \hspace{-1mm}
    \includegraphics[width=0.9\linewidth]{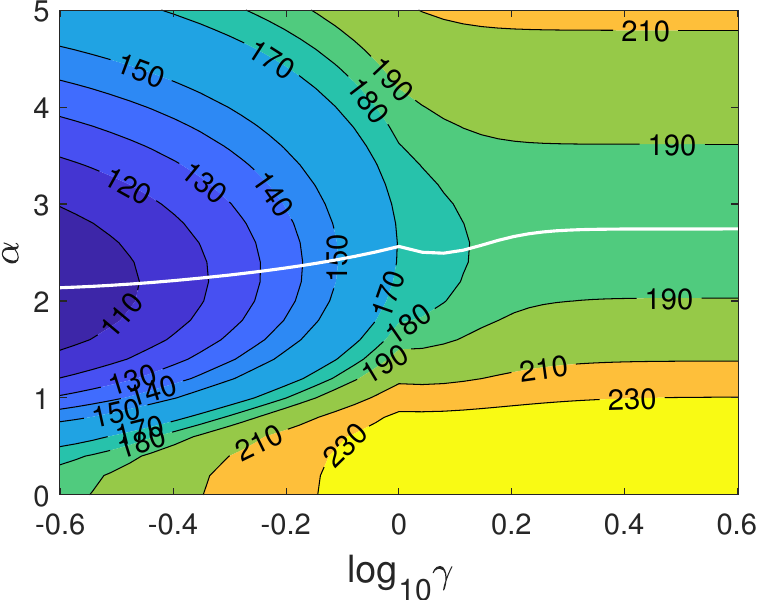}
    \phantomsubcaption
    \label{fig:ductHa0_c}
  \end{subfigure}
  \hfill
  \begin{subfigure}[b]{0.49\textwidth}
    \footnotesize (d) \hspace{-1mm}
    \includegraphics[width=0.9\linewidth]{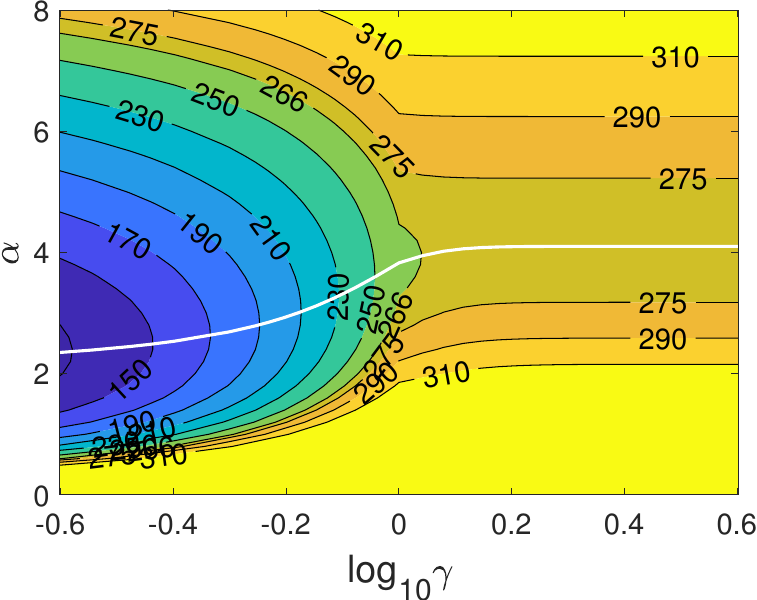}
    \phantomsubcaption
    \label{fig:ductHa0_d}
  \end{subfigure}
  \caption{Cartography of $\Rey_E$ in the $(\alpha,\mbr{\log_{10}}(\gamma))$ plane. Duct  geometry, from (\subref{fig:ductHa0_a}) to (\subref{fig:ductHa0_d}) $\Ha=0,5,10,20$.
  }
  \label{fig:ductHa0}
\end{figure}

\begin{table}
  \begin{center}
    \begin{tabular*}{.5\textwidth}{@{\extracolsep{\fill}}cccc@{}}
      {$\Ha$} & $\gamma$ & {$N_y$} & {$N_z$} \\[3pt]
      $0$ & $\ge 1$  & $\left[27\gamma\right]-2 $ &  $25$ \\
      $5$ & $ \ge 1$ & $\left[31\gamma\right]-2$ &  $33$ \\
      $10$ & $ \ge 1$ & $\left[32\gamma\right]-2$ &  $36$ \\
      $20$ & $ \ge 1$ & $\left[42\gamma\right]-2$ &  $45$ \\[2pt]
      $5$ & $< 1$ & $28$ & $\left[38/\gamma\right]-2$  \\
      $10$ & $< 1$ & $30$ & $\left[38/\gamma\right]-2$  \\
      $20$ & $< 1$ & $40$ & $\left[47/\gamma\right]-2$  \\
    \end{tabular*}
  \end{center}
  \caption{Resolutions for the computations of figure \ref{fig:ductHa0} indicated by maximum order of Chebyshev polynomials. The square brackets denote the integer part.  }
  \label{table:resolutions}
\end{table}

\subsection{Energy stability for $\Ha=0$}
\label{sec:energy_stability_ha0}

Figure \ref{fig:ductHa0} shows color maps of the values of $\Rey_E$ in an $(\alpha,\gamma)$ plane, where $\alpha$ is the streamwise wavenumber and $\gamma$ is represented in (base 10) logarithmic scale.
Values of $\Ha=0,5,10$ and $20$ are shown. The numerical resolutions for these computations are given in table \ref{table:resolutions}.

We focus first on figure \ref{fig:ductHa0_a} which has $\Ha=0$. As far as we know no exhaustive energy stability study has been performed in rectangular duct flow even in the absence of MHD effects. This configuration, where $\Ha=0$, is characterised by an additional degree of symmetry compared to the MHD case: all sidewalls are equivalent and the notion of Shercliff and Hartmann walls is irrelevant. Mathematically this results in the equivalence between an aspect ratios $\gamma=L_y/L_z>0$ and its inverse $1/\gamma=L_z/L_y$. We thus expect the symmetric relation $\Rey_E(\gamma)=\Rey_E(1/\gamma)$ to be valid. This should manifest itself graphically in a flip symmetry with respect to the zero axis when plots are made according to the variable $\log{\gamma}$. As expected, the symmetry property $\Rey_E(\gamma)=\Rey_E(1/\gamma)$ is clearly visible in the figure.

The location of the wavenumber $\alpha=\alpha_m$ associated with the optimal value $\Rey_E$ has been represented in Figure \ref{fig:ductHa0} for each value of $\gamma$, by using a plain white line. For the case $\Ha=0$, non-zero values of $\alpha_m$ appear to be restricted to an interval where $|\log_{10}(\gamma)| \lesssim 0.2 $, \ie $0.6 \lesssim \gamma \lesssim 1.6$. The largest value of $\alpha_m$ (\ie the shortest wavelength) is found on the symmetry axis for $L_y=L_z$ and corresponds to the cusp in the figure. Outside this interval the wavenumber minimising $\Rey_E$ is everywhere zero.

The variation of $\Rey_E$ (at optimum wavenumber) with $\gamma$ is shown in figure \ref{fig:ReEvsgamma_a}. As one would expect, the values for $\Ha=0$ approach the channel limit with $\Rey=49.6$ for decreasing as well as for increasing $\gamma$.  One can also notice two discontinuities in the slope of the curve $\Rey_E(\gamma)$ on either side of $\gamma=1$. For $\gamma >1$, these occur at $\gamma\approx 1.8$ and $\gamma\approx 2.8$. They correspond to a qualitative change in the structure of the mode providing $\Rey_E$, which will be shown in \S\ref{sec:structures}.

We focus now on the square case, \ie $\gamma=1$. In the literature, to our knowledge only a numerical value of $\Rey_E = 79.44$ (based on the centerline velocity) has been reported by \cite{biau2008transition} in the absence of MHD effects, associated with a zero streamwise wavenumber. This is at odds with our result $\alpha = 1.3$ corresponding to a smaller value $\Rey_E = 74.1$. For $\alpha=0$ we obtain $\Rey_E=78.5$ in reasonable agreement with \cite{biau2008transition}, where  a different numerical method was used.
We note for comparison that the companion circular geometry of Hagen-Poiseuille flow also features a non-zero axial optimal wavenumber $\alpha=1.07$ found for $\Rey_E = 81.5$ \citep{joseph1969stability}.

\subsection{Influence of increasing $\Ha$}

As $\Ha$ increases above zero, the flip symmetry in figure \ref{fig:ductHa0_a} is immediately lost. This corresponds to an increasing dissymmetry between
the two different pairs of boundary layers along the side wall: the Hartmann and the Shercliff boundary layers are now two distinct boundary layers with different scalings. The minimal value of $\Rey_E$ in figure \ref{fig:ductHa0} is always achieved, unlike for $\Ha = 0$, for a finite wavenumber $\alpha_m$. This minimum is always found at the lowest computed $\gamma$ values.
This corresponds to the configuration where the longer edge is parallel to the magnetic field: the laminar base flow is then dominated by wider Shercliff layers and the two thinner Hartmann layers are well separated. The minimal value of $\Rey_E$ itself increases with $\Ha$. For $\Ha=0,5,10,20$, it is respectively 52.2, 88.1, 102.2 and 127.3.

\begin{figure}
  \centering
  \begin{subfigure}[b]{0.49\textwidth}
    \setlength{\unitlength}{1.0cm}
    %\fbox{
    \begin{picture}(6.0,4.5)
      \put(0.0,0.0){\includegraphics[height=0.68\linewidth]{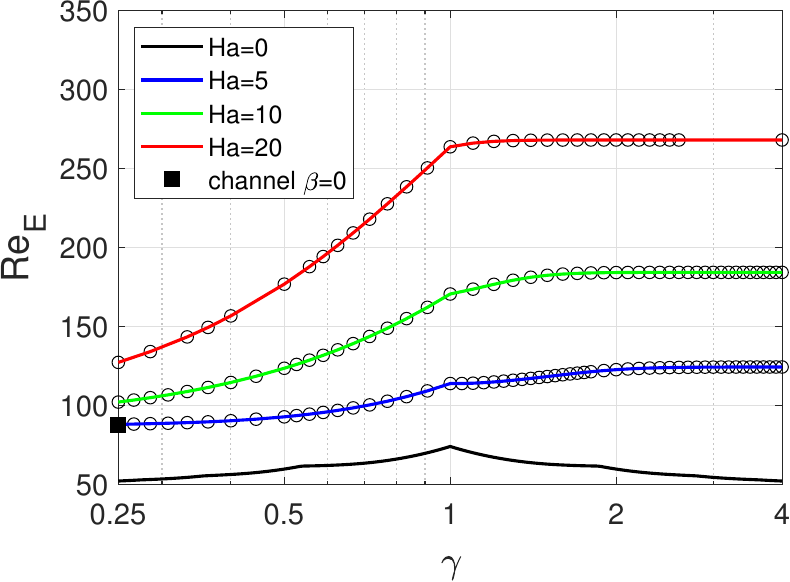}}%
      \put(0.0,0.0){\footnotesize (a)}
    \end{picture}
    %}
    \phantomsubcaption
    \label{fig:ReEvsgamma_a}
  \end{subfigure}
  %\hfill
  \begin{subfigure}[b]{0.49\textwidth}
    \setlength{\unitlength}{1.0cm}
    %\fbox{
    \begin{picture}(6.0,4.5)
      \put(0.4,0.0){\includegraphics[height=0.68\linewidth]{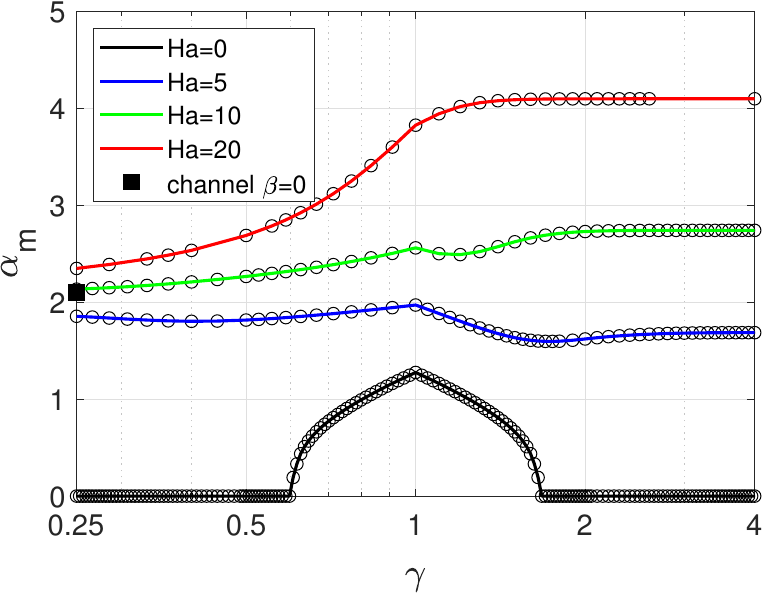}}%
      \put(0.0,0.0){\footnotesize (b)}
    \end{picture}
    %}
    \phantomsubcaption
    \label{fig:ReEvsgamma_b}
  \end{subfigure}
  \caption{$\Rey_E$ and $\alpha_m$ vs.~$\gamma$ in the duct geometry.%{\color{cyan}MBR: I suppose that the circles represent our calculations. Should we not mark these for $\Ha=0$ also?}
  }
  \label{fig:ReEvsgamma}
\end{figure}

The global trend for the value of $\Rey_E$ is an increase with $\Ha$, which is also seen in figure \ref{fig:ReEvsgamma_a}. It appears that $\Rey_E$ and the corresponding wavenumber $\alpha$ shown in figure \ref{fig:ReEvsgamma_b} approach  Orr's value represented by a black square on the left axis $\gamma=0.25$ for all $\Ha\ge 5$. This is consistent with the channel flow with  spanwise magnetic field because the base flow in the duct approaches the Poiseuille profile as $\gamma$ tends to zero (with exception of the Hartmann layers).

Figures \ref{fig:ReEvsgamma_a} and \ref{fig:ReEvsgamma_b} also show that a plateau emerges for both $\Rey_E$ and $\alpha_m$ at large $\gamma$. The higher the value of $\Ha$, the earlier the plateau is reached as $\gamma$ is increased.
For $\Ha \ge 5$, $\alpha_m$ stays away from zero for all aspect ratios $\gamma$ shown. The range of values of $\alpha_m$ found by varying $\gamma$ shifts upwards as $\Ha$ is increased. This corresponds, as $\Ha$ gets larger, to increasingly shorter wavelengths found at criticality. For $\Ha=10$ and beyond, the shorter wavelengths are found for $\gamma>1$.

\subsection{Connection with the Quasi Two-dimensional theory}

\begin{figure}
  \centering
  \begin{subfigure}[b]{0.49\textwidth}
    \setlength{\unitlength}{1.0cm}
    %\fbox{
    \begin{picture}(6.0,4.5)
      \put(0.0,0.0){\includegraphics[height=0.68\linewidth]{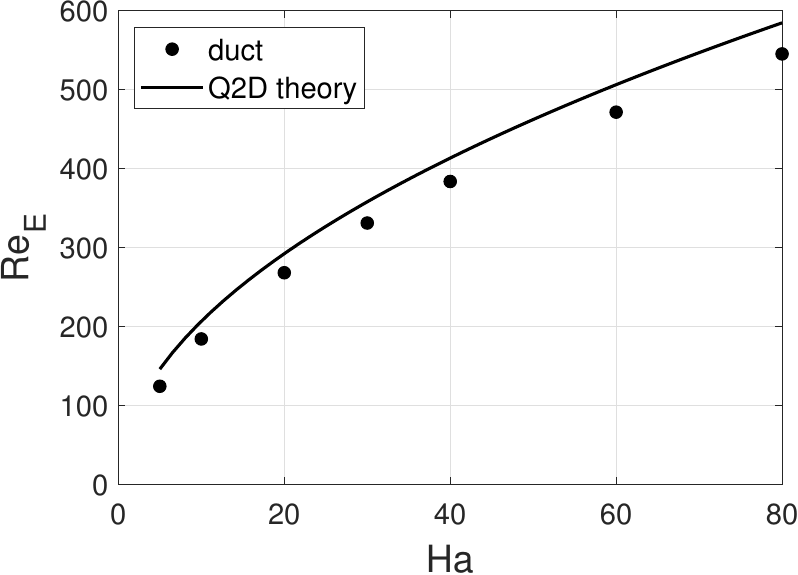}}%
      \put(0.0,0.0){\footnotesize (a)}
    \end{picture}
    %}
    \phantomsubcaption
    \label{fig:q2dcomp_a}
  \end{subfigure}
  %\hfill
  \begin{subfigure}[b]{0.49\textwidth}
    \setlength{\unitlength}{1.0cm}
    %\fbox{
    \begin{picture}(6.0,4.5)
      \put(0.39,0.03){\includegraphics[height=0.665\linewidth]{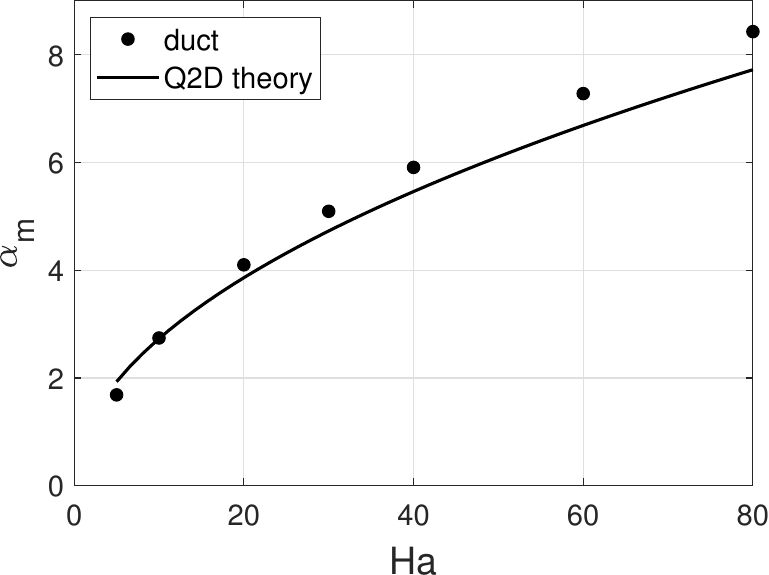}}%
      \put(0.0,0.0){\footnotesize (b)}
    \end{picture}
    %}
    \phantomsubcaption
    \label{fig:q2dcomp_b}
  \end{subfigure}
  \caption{\yd{(a) $\Rey_E$ vs. ~$\Ha$, (b) $\alpha_m$ vs.~$\Ha$} in the duct geometry (in the large $\gamma$ limit), with comparison with equation \eqref{eq:potherat} from Q2D theory.}
  \label{fig:q2dcomp}
\end{figure}

We investigate now the other limiting configuration $\gamma \rightarrow \infty$ where the shorter edge is parallel to the magnetic field.
This is associated visually with the right of each subplot in figure \ref{fig:ductHa0}, in which the same values of $\Ha=0,5,10$ and $20$ are displayed.
In this configuration, the laminar flow consists of two narrow Shercliff layers and two laterally extended Hartmann layers.
From figure \ref{fig:ductHa0} it is clear that, at least for $\Ha \neq 0$, $\Rey_E$ achieves for asymptotically large $\gamma$ a minimum value associated with non-zero values of $\alpha_m$. The corresponding values of $\Rey_E$ and $\alpha_m$ are reported in figure \ref{fig:q2dcomp}, respectively \subref{fig:q2dcomp_a} and \subref{fig:q2dcomp_b}.
The additional values of $\Rey_E$ and $\alpha_m$ for $\Ha>20$ were typically computed at  two  distinct values of $\gamma>1$. This was done in order to ensure that the plateau is reached without going to the computationally expensive case $\gamma=4$.

Both $\Rey_E$ and $\alpha_m$ increase monotonically with increasing $\Ha$. This is interpreted, for this large $\gamma$ limit, as a delay of the transition by the magnetic field, associated with smaller axial wavelengths at criticality.

\begin{figure}
  \centering
  \begin{subfigure}[b]{0.49\textwidth}
    \setlength{\unitlength}{1.0cm}
    %\fbox{
    \begin{picture}(6.0,4.5)
      \put(0.1,0.0){\includegraphics[height=0.68\linewidth]{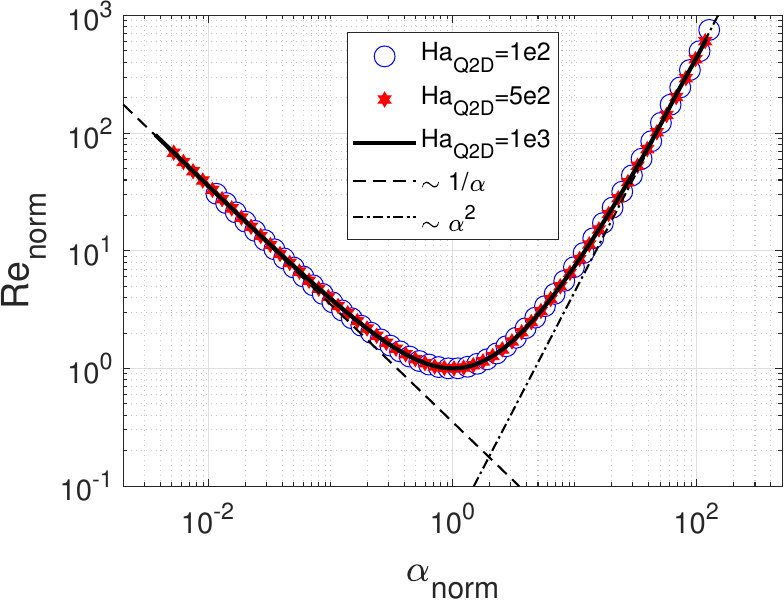}}%
      \put(0.0,0.0){\footnotesize (a)}
    \end{picture}
    %}
    \phantomsubcaption
    \label{fig:q2dcomp2_a}
  \end{subfigure}
  %\hfill
  \begin{subfigure}[b]{0.49\textwidth}
    \setlength{\unitlength}{1.0cm}
    %\fbox{
    \begin{picture}(6.0,4.5)
      \put(0.2,0.08){\includegraphics[height=0.662\linewidth]{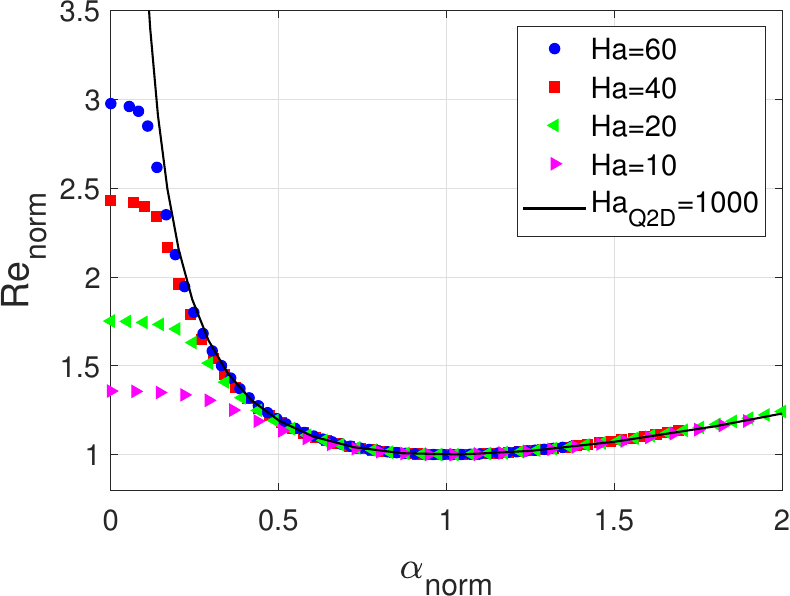}}%
      \put(0.0,0.0){\footnotesize (b)}
    \end{picture}
    %}
    \phantomsubcaption
    \label{fig:q2dcomp2_b}
  \end{subfigure}
  \caption{$\Rey_E$ vs.~$\alpha$, with all quantities rescaled by their value at the minimum $\Rey_E$. (\subref{fig:q2dcomp2_a})
  Rescaled results from the Q2D model for several $\Ha_{Q2D}$. (\subref{fig:q2dcomp2_b})
  Comparison of duct results (in the limit $\gamma \gg 1$) with results from the Q2D model.}
  \label{fig:q2dcomp2}
\end{figure}

The present results can be compared with earlier work \citep{potherat2007quasi} carried out in the framework of the quasi two-dimensional (Q2D) approximation \citep{sommeria1982and,potherat2000effective}.
In the Q2D model, the flow is represented by a  two-dimensional  velocity field that corresponds to the actual flow averaged along the direction of the magnetic field. The averaged flow satisfies the two-dimensional Navier-Stokes equations with an additional  linear damping term  $-\Ha_{Q2D} \bm{u}$. This term accounts for the friction on the Hartmann walls. The Q2D Reynolds number is defined with the lateral dimension $L_y/2$, \ie $ \Rey_{Q2D} = \gamma \Rey$. Correspondingly, $\alpha_{Q2D}=\alpha/\gamma$. The relation between $\Ha_{Q2D}$ and $\Ha$ is $\Ha_{Q2D}= \gamma^2 \Ha/2$. The basic flow in the Q2D model is equivalent to the Hartmann flow profile but with a side layer thickness $\sim \Ha_{Q2D}^{-1/2}$.
For $\Ha_{Q2D} \gg 1$, the Q2D energy stability analysis shows a universal, self-similar dependence between Reynolds and wavenumber illustrated in figure \ref{fig:q2dcomp2_a}.
This agreement between different $\Ha_{Q2D}$ demonstrates that the (averaged) Shercliff layers on the opposite walls become decoupled, \ie outer length scale (width of the duct) does not affect the result. In particular, the minimum  $\alpha_{Q2D}$ and the corresponding $\Rey_{Q2D}$ therefore scale as $\Ha_{Q2D}^{1/2}$. The numerical values of the coefficients in the scaling relations are given in \cite{potherat2007quasi}. Transformed to our definitions, they read
\begin{subequations}
  \label{eq:potherat}
  \begin{gather}
    \Rey= 65.3 \sqrt{\Ha}, \qquad
    \alpha=0.863\sqrt{\Ha}.
    \tag{\theequation a,b}
  \end{gather}
\end{subequations}
The qualitative as well as quantitative match between the present computations and the Q2D results in figure \ref{fig:q2dcomp} is good, despite a slight drift observed for the largest values of $\Ha$ (computed here up to $Ha=80$). The perturbations are therefore expected to become localized in the Shercliff layers, and to exhibit a shape with approximate uniformity along the magnetic field.

We can further verify the universal scaling behavior for our duct results over intervals of $\alpha$. This is reported in figure \ref{fig:q2dcomp2_b}. It shows the dependence of normalized $\Rey_E$ and $\alpha$ for different $\Ha$ as well as the universal curve from the Q2D model in figure \ref{fig:q2dcomp2_a}. The agreement with the Q2D model is excellent except in the limit $\alpha\to 0$, where the duct curves depart from the Q2D theory. In contrast to the Q2D asymptotic behavior $\Rey\sim 1/\alpha$, they saturate at finite values of $\Rey$ for $\alpha=0$. These values increase monotonically with $\Ha$.

As already noted, the dependence  $\Rey(\alpha)$ in the Q2D model appears to be a power law for $\alpha\to 0$ and for $\alpha\to \infty$ (see figure \ref{fig:q2dcomp2_a}). The former is consistent with a regular limit $\alpha\to 0$ in the Q2D equations.  The $O(\alpha^2)$ scaling for large $\alpha$ cannot be justified in this way. It can be understood as a diffusive scaling associated with wavelengths with a viscous damping rate $O(\alpha^2\Rey^{-1})$ so high that it requires $\Rey=O(\alpha^2)$ for damping to be balanced instantaneously by non-normal amplification.

\subsection{Case $\alpha=0$}
\label{sec:alphazero}
\begin{figure}
  \centering
  \begin{subfigure}[h]{.49\textwidth}
    \setlength{\unitlength}{1.0cm}
    %\fbox{
    \begin{picture}(6.0,4.5)
      \put(0.0,0.0){\includegraphics[height=0.68\linewidth]{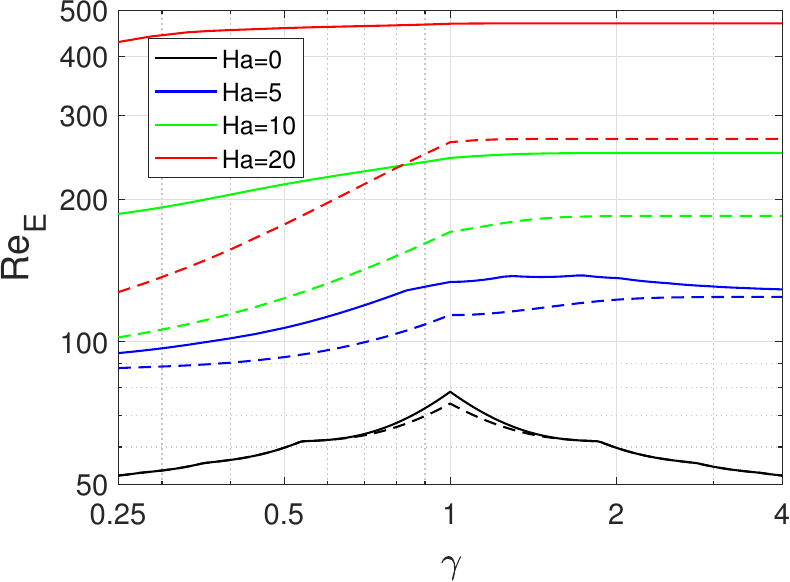}}%
      \put(0.0,0.0){\footnotesize (a)}
    \end{picture}
    %}
    \phantomsubcaption
    \label{fig:ductHaalpha0_a}
  \end{subfigure}
  %\hfill
  \begin{subfigure}[h]{0.49\textwidth}
    \setlength{\unitlength}{1.0cm}
    %\fbox{
    \begin{picture}(6.0,4.5)
      \put(0.0,0.04){\includegraphics[height=0.664\linewidth]{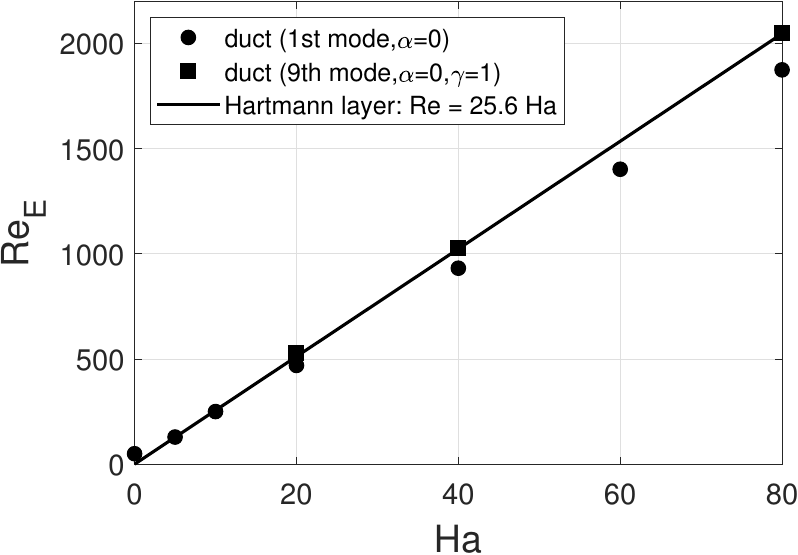}}%
      \put(0.0,0.0){\footnotesize (b)}
    \end{picture}
    %}
    \phantomsubcaption
    \label{fig:ductHaalpha0_b}
  \end{subfigure}
  \caption{(\subref{fig:ductHaalpha0_a}) $\Rey_E$ for $\alpha=0$ (full lines) and optimal $\alpha$ (dashed) vs.~$\gamma$ in duct  geometry.
  (\subref{fig:ductHaalpha0_b}) $\Rey_E$ for $\alpha=0$ (duct, limit $\gamma\gg 1$) vs.~$\Ha$ and comparison with Hartmann layer scaling.}
  \label{fig:ductHaalpha0}
\end{figure}

While the disturbances with $\alpha=0$ do not minimise $\Rey_E$ for $\Ha\ge 5$, they are interesting in their own right since their corresponding $\Rey_E$ has a different dependence on $\Ha$ than the optimal mode. This is apparent from figure
\ref{fig:ductHaalpha0_a} that shows the $\gamma$-dependence of $\Rey_E$ for $\alpha=\alpha_m$ and $\alpha=0$. For large $\gamma$, $\Rey_E$ for $\alpha=0$ saturates like the minimal $\Rey_E$ but the saturation levels for $\alpha_m$ and $\alpha=0$ separate further as $\Ha$ grows.
For small $\gamma$, a saturation is only apparent for $\Ha=5$ with comparable levels for $\alpha_m$ and $\alpha=0$. For $\Ha= 10$  and $\Ha=20$ the curves for  $\alpha_m$ and $\alpha=0$ continue to decay below $\gamma=0.25$. It can be expected that the curves for $\alpha=0$ eventually reach the saturation levels for the channel case. These  correspond  to the minima of $\Rey_E$ on the axis $\alpha=0$ in figures \ref{fig:channelHa0_c}, \ref{fig:channelHa0_d}. For $\Ha=10$ and $\Ha=20$ these values are $\Rey_E\approx 158$ and  $\Rey_E\approx 310$, respectively.

The saturated values of $\Rey_E$ for $\alpha=0$ and $\gamma>1$ are shown in figure \ref{fig:ductHaalpha0_b} as black circles. They clearly scale as $\Rey_E\sim O(\Ha)$ with a proportionality constant $\approx 23$.
This linear scaling is consistent with the behavior of linear, streamwise-independent optimal perturbations investigated by \cite{krasnov2010optimal}. These authors found that those perturbations  reside in the Shercliff layers. A linear scaling was also obtained by \cite{lingwood1999stability}. These authors investigated energy stability for a single insulating Hartmann layer. They found $\Rey_E=25.6\, \Ha$ for purely streamwise-independent modes ($\alpha=0$). Although the numerical value of the proportionality constant for the duct is close to $25.6$, the corresponding modes are distinct. Perturbations localized in the Hartmann layers only appear as higher modes in the duct case. For $\gamma=1$, the 9th mode is the lowest one with such a spatial structure. The corresponding $\Rey_E$ of the 9th mode (also displayed in figure \ref{fig:ductHaalpha0_b}) agrees very well with the energy stability limit from \cite{lingwood1999stability}.

In summary, the case $\alpha=0$ provides a linear dependence between Reynolds and Hartmann numbers for energy stability. This is consistent with threshold values of $\Rey$ found in experiments on the relaminarization of turbulent MHD duct flows \citep{branover1978magnetohydrodynamic,moresco2004experimental}. A scaling with $\Ha^{1/2}$ is not observed in those experiments. This underlines the dynamical importance of long-wave  modes in transitional and turbulent MHD duct flows despite their non-optimal properties.

\section{Coherent structures at criticality}
\label{sec:structures}

\subsection{Theoretical link with linear optimal perturbations (LOPs)}

Although many energy stability calculations principally report values of $\Rey_E$ and the corresponding optimal wavenumbers at criticality (\ie at $\Rey=\Rey_E$), the associated flow structures, corresponding to the eigenvectors of the linearised operator in \eqref{eq:energystability2}, are usually not investigated in detail with the possible exception of \cite{potherat2000effective}. These coherent structures are not directly observable flow structures in an experimental setting, unlike \eg unstable global modes. However, as mentioned earlier, they bear a strong relation with the linear optimal perturbations (LOPs) celebrated in non-modal instability analysis \citep{schmid2007nonmodal,kerswell2018nonlinear}. The LOPs are the initial velocity fields ${\bm u}^{(T)}_{\mathrm{opt}}$ that optimise the finite-time perturbation energy growth
\begin{equation}
    G(T,{\bm u}(0))=|{\bm u}(T)|^2/|{\bm u}(0)|^2
\end{equation}
for any given time horizon $T>0$, under the action of the linearised dynamics \citep{reddy1993energy}. They are relevant for $\Rey>\Rey_E$, when energy growth is indeed instantaneously possible and $\max{(G)}>1$ for at least some value of $T$. The maximum energy growth corresponding to the short time horizons $\Delta t \ll 1$ verifies $G(\Delta t,{\bm u}^{\Delta t}_{\mathrm{opt}}) \le \lVert \bm{\mathcal{I}}+2(\Delta t)\bm{\mathcal{L}} \rVert$ at first order in $\Delta t$, where $\bm{\mathcal{L}}$ is the linearised operator at $t=0$. This tends to unity, for fixed $\Delta t$, as $\Rey$ approaches $\Rey_E$ from above. The structures achieving the largest energy growth at $\Rey=\Rey_E$ are hence precisely the critical perturbations computed in all energy stability studies as a byproduct of the eigenvalue problem  \eqref{eq:energystability2}. In other words, the LOPs computable at $\Rey>\Rey_E$ continue smoothly into the critical perturbations computed here. In the same way as the detailed investigation of the LOPs has shed light on the (linear) transition mechanisms, it is hence useful to investigate the critical perturbations at $\Rey_E$ as they might already feature elements related to the transition observed at larger $\Rey$.
This is for instance the case for the channel flow discussed in \S\ref{sec:channel_results}, where the mode that attains the lowest $\Rey_E$ changes from a streamwise to a spanwise uniform structure as $\Ha$ is increased, and for intermediate parameter values correspond to an oblique wave.
Such a trend is in complete agreement with the behaviour of LOPs reported by \cite{krasnov2008optimal} (see their figure 7).
On a technical level, the critical perturbations computed from \eqref{eq:energystability2} are independent of any target time, which makes their description simpler.

It is useful to recall the main teachings of the quest for LOPs in simple configurations, namely the purely hydrodynamic channel flow with streamwise periodicity. Two-dimensional computations (assuming no spanwise dependence of the flow) have highlighted the Orr mechanism as the most efficient way to extract energy from the base flow \citep{farrell1988optimal}. The Orr mechanism consists of a progressive shearing of the perturbations in the direction associated with the base flow. The corresponding optimal perturbations are easily recognised by the tilting of spanwise vortices \textit{against} the shear. Three-dimensional computations of LOPs have highlighted a much more efficient energy growth mechanism, linked to the \textit{lift-up} mechanism \citep{brandt2014lift} which actively exploits the spanwise dependence of the disturbance. The associated optimal perturbations look like long tubular streamwise vortices evolving rapidly into streamwise streaks, characterised by a well-defined spanwise spacing. These optimal disturbances are often two-dimensional \citep{butler1992three}, now in the sense that they do not depend on the streamwise coordinate, and even when they are three-dimensional, the tilting of the vortices against the shear is not pronounced.

\subsection{Duct visualisation for $\Ha=0$.}

\begin{figure}
  \centering
  \begin{subfigure}[b]{0.32\textwidth}
    \footnotesize (a) \hspace{-1mm}
    \includegraphics[width=0.9\linewidth]{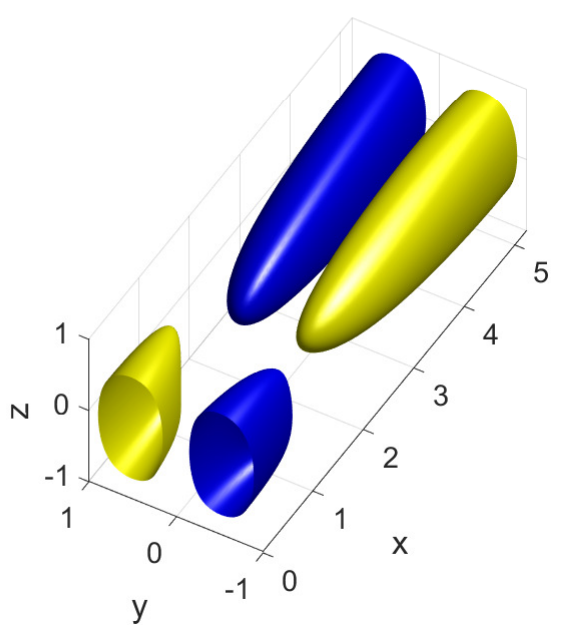}
    \phantomsubcaption
    \label{fig:ductHa03d_a}
  \end{subfigure}
  \begin{subfigure}[b]{0.32\textwidth}
    \footnotesize (b) \hspace{-1mm}
    \includegraphics[width=0.9\linewidth]{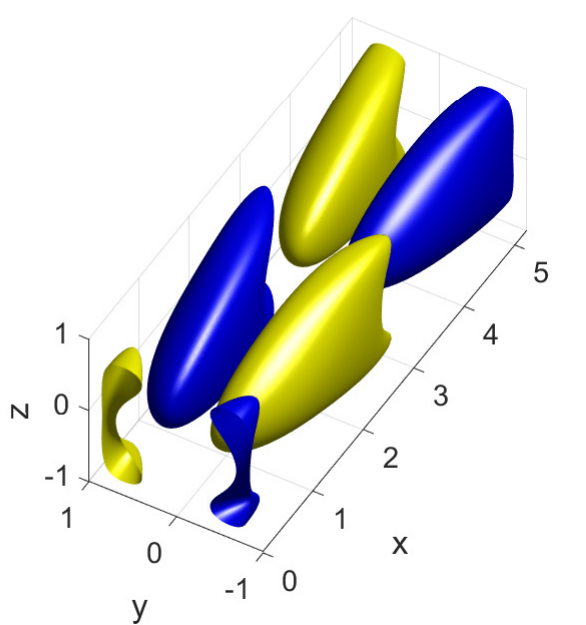}
    \phantomsubcaption
    \label{fig:ductHa03d_b}
  \end{subfigure}
  \begin{subfigure}[b]{0.32\textwidth}
    \footnotesize (c) \hspace{-1mm}
    \includegraphics[width=0.9\linewidth]{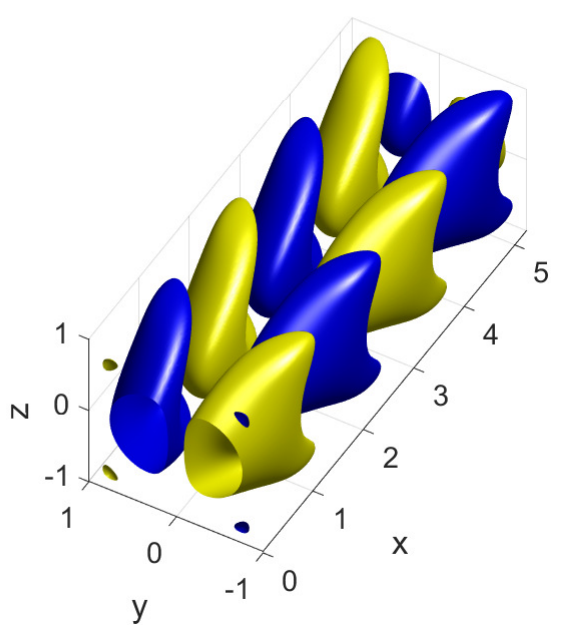}
    \phantomsubcaption
    \label{fig:ductHa03d_c}
  \end{subfigure}
  \caption{Isosurfaces of $u_x$ for the leading eigenmodes at $\Ha=0$, $\gamma=1$ and streamwise wavenumbers (\subref{fig:ductHa03d_a}) $\alpha=0.6$, (\subref{fig:ductHa03d_b}) $\alpha=1.2$ and (\subref{fig:ductHa03d_c}) $\alpha=2.4$.
  %In the case $\alpha=0.6$, only half the  streamwise period is shown.
  \yd{The streamwise period displayed in each case is the period for $\alpha=1.2$.}
  }
  \label{fig:ductHa03d}
\end{figure}

\begin{figure}
  \centering
  \begin{subfigure}[b]{0.32\textwidth}
    \footnotesize (a) \hspace{-1mm}
    \includegraphics[width=0.9\linewidth]{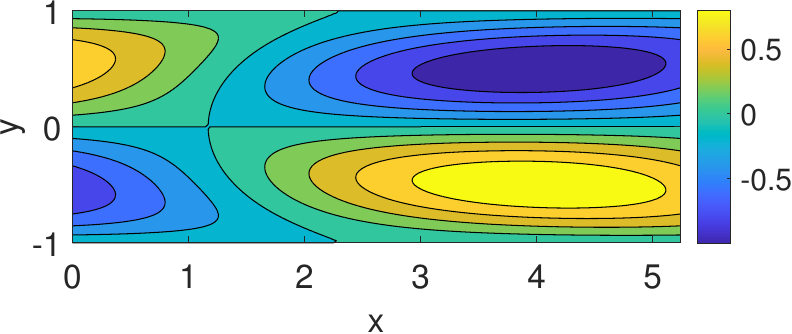}
    \phantomsubcaption
    \label{fig:ductHa03d2d_a}
  \end{subfigure}
  \begin{subfigure}[b]{0.32\textwidth}
    \footnotesize (b) \hspace{-1mm}
    \includegraphics[width=0.9\linewidth]{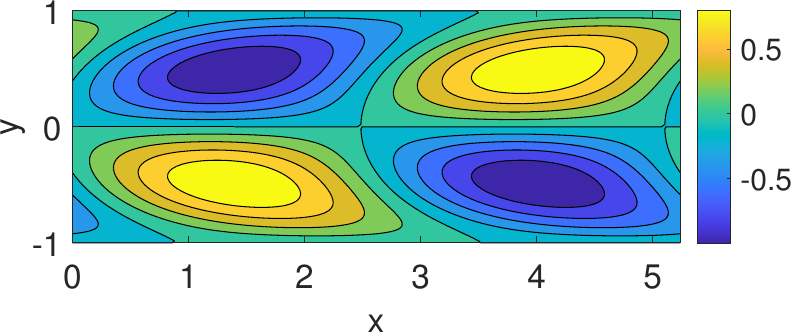}
    \phantomsubcaption
    \label{fig:ductHa03d2d_b}
  \end{subfigure}
  \begin{subfigure}[b]{0.32\textwidth}
    \footnotesize (c) \hspace{-1mm}
    \includegraphics[width=0.9\linewidth]{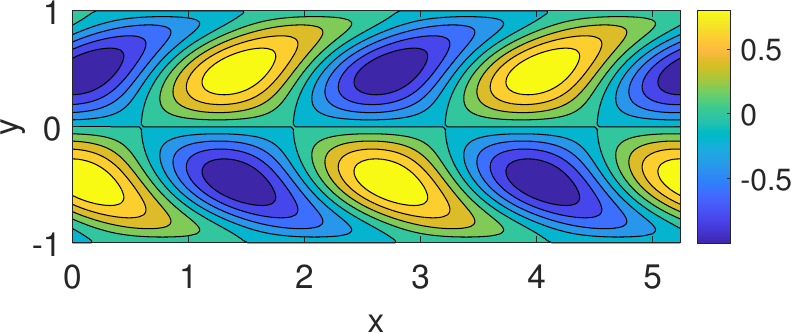}
    \phantomsubcaption
    \label{fig:ductHa03d2d_c}
  \end{subfigure}
    \begin{subfigure}[b]{0.32\textwidth}
    \footnotesize (d) \hspace{-1mm}
    \includegraphics[width=0.9\linewidth]{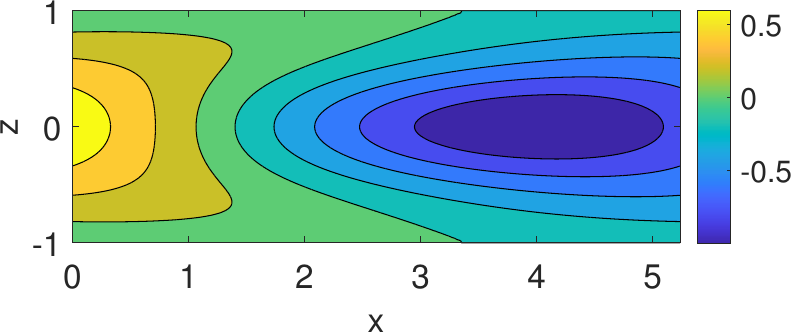}
    \phantomsubcaption
    \label{fig:ductHa03d2d_d}
  \end{subfigure}
  \begin{subfigure}[b]{0.32\textwidth}
    \footnotesize (e) \hspace{-1mm}
    \includegraphics[width=0.9\linewidth]{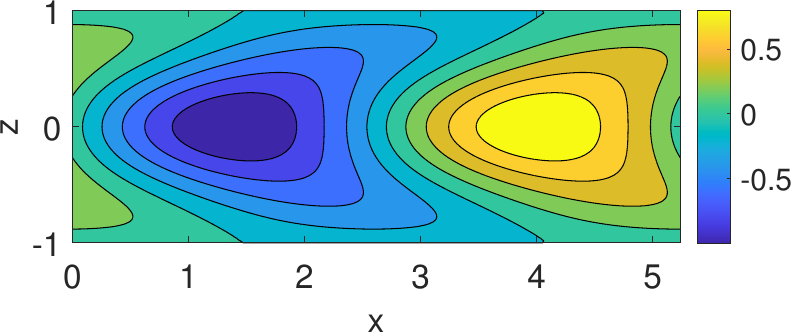}
    \phantomsubcaption
    \label{fig:ductHa03d2d_e}
  \end{subfigure}
  \begin{subfigure}[b]{0.32\textwidth}
    \footnotesize (f) \hspace{-1mm}
    \includegraphics[width=0.9\linewidth]{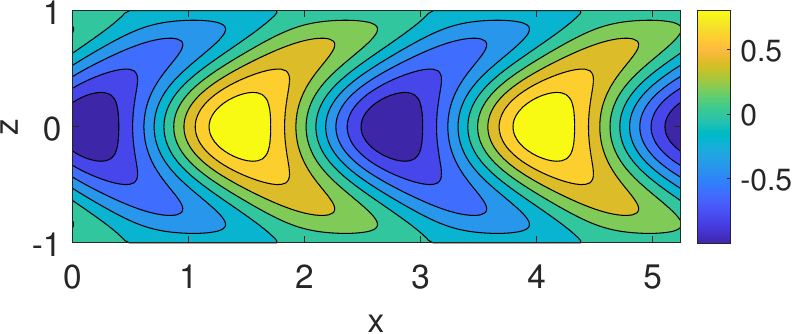}
    \phantomsubcaption
    \label{fig:ductHa03d2d_f}
  \end{subfigure}
  \caption{Contours of $u_x$ in the planes $z=0$ (\subref{fig:ductHa03d2d_a}-\subref{fig:ductHa03d2d_c}) and $y=1/2$ (\subref{fig:ductHa03d2d_d}-\subref{fig:ductHa03d2d_f}) for the leading eigenmodes at $\Ha=0$ for the same parameters as Figure \ref{fig:ductHa03d}.
%  , $\gamma=1$ and streamwise wavenumbers (\subref{fig:ductHa03d2d_a},\subref{fig:ductHa03d2d_d}) $\alpha=0.6$, (\subref{fig:ductHa03d2d_b},\subref{fig:ductHa03d2d_e}) $\alpha=1.2$ and (\subref{fig:ductHa03d2d_c},\subref{fig:ductHa03d2d_f}) $\alpha=2.4$.
  %In the case $\alpha=0.6$, only half the  streamwise period is shown.
  %{\color{cyan}MBR: (\subref{fig:ductHa03d2d_d}-\subref{fig:ductHa03d2d_f}) should perhaps have $z$ as the label on the vertical axis?}
  }
  \label{fig:ductHa03d2D}
\end{figure}

We begin by visualising the eigenmodes of the eigenproblem \eqref{eq:energystability2} in the non-MHD case when $\Ha=0$. Starting with the least stable modes and focusing on the square duct ($\gamma=1$), figure \ref{fig:ductHa03d} shows three-dimensional rendering of the isosurfaces of $u_x$  for three values of $\alpha$ from low to high, respectively $\alpha=0.6,1.2$ and $2.4$.
\yd{The tilting against the shear typical of Orr modes is found visually in figure \ref{fig:ductHa03d}. This tilting is confirmed by looking at the perturbations both in the $xy$ and $xz$ planes, as shown in figure \ref{fig:ductHa03d2D} for the same values of $\alpha$.}

%This suggests, as in channel flow, that long modes with small or vanishing $\alpha$ are linked at higher $\Rey$ with the lift-up mechanism while the Orr mechanism is present in shorter-wavelength structures.

\begin{figure}
  \centering
  \begin{subfigure}[b]{0.49\textwidth}
    \footnotesize (a) \hspace{-1mm}
    \includegraphics[width=0.9\linewidth]{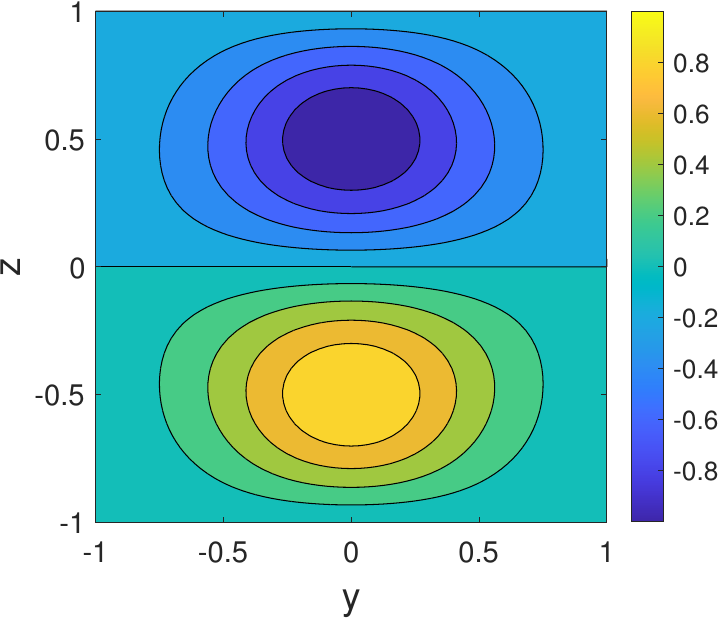}
    \phantomsubcaption
    \label{fig:ductHa0a03_a}
  \end{subfigure}
  \hfill
  \begin{subfigure}[b]{0.49\textwidth}
    \footnotesize (b) \hspace{-1mm}
    \includegraphics[width=0.9\linewidth]{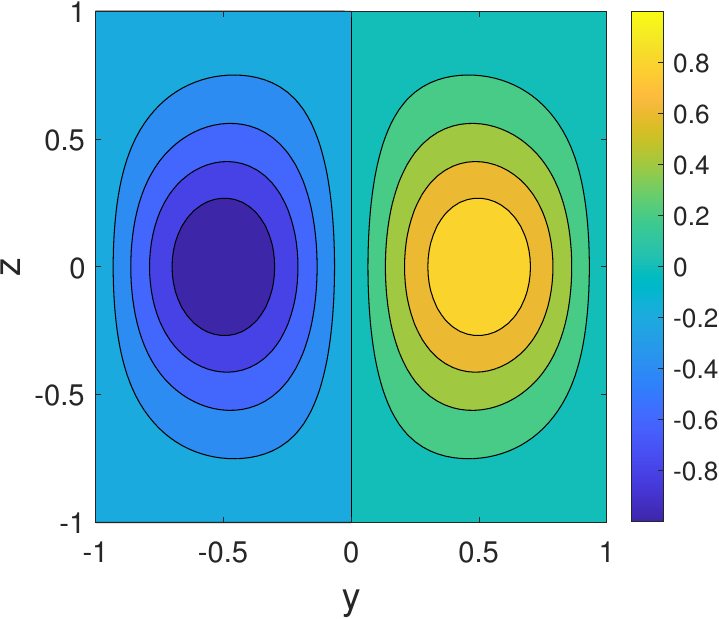}
    \phantomsubcaption
    \label{fig:ductHa0a03_b}
  \end{subfigure}\\
  \begin{subfigure}[b]{0.49\textwidth}
    \footnotesize (c) \hspace{-1mm}
    \includegraphics[width=0.9\linewidth]{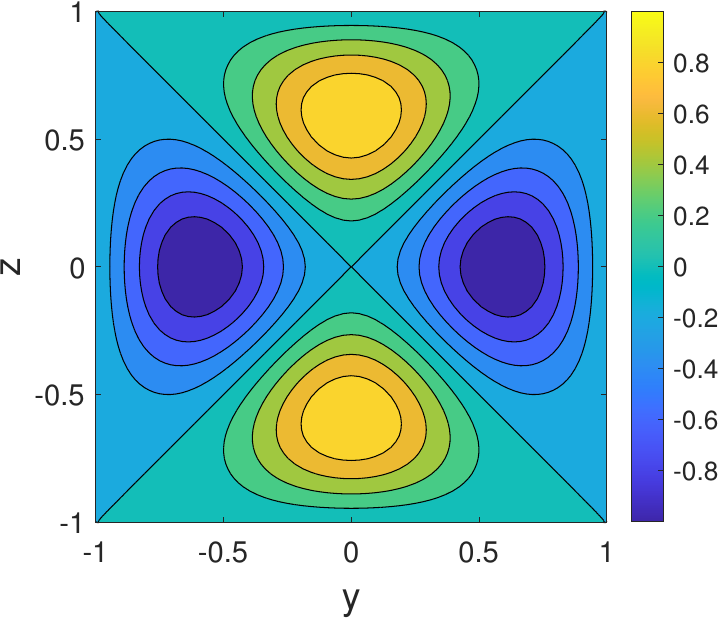}
    \phantomsubcaption
    \label{fig:ductHa0a03_c}
  \end{subfigure}
  \hfill
  \begin{subfigure}[b]{0.49\textwidth}
    \footnotesize (d) \hspace{-1mm}
    \includegraphics[width=0.9\linewidth]{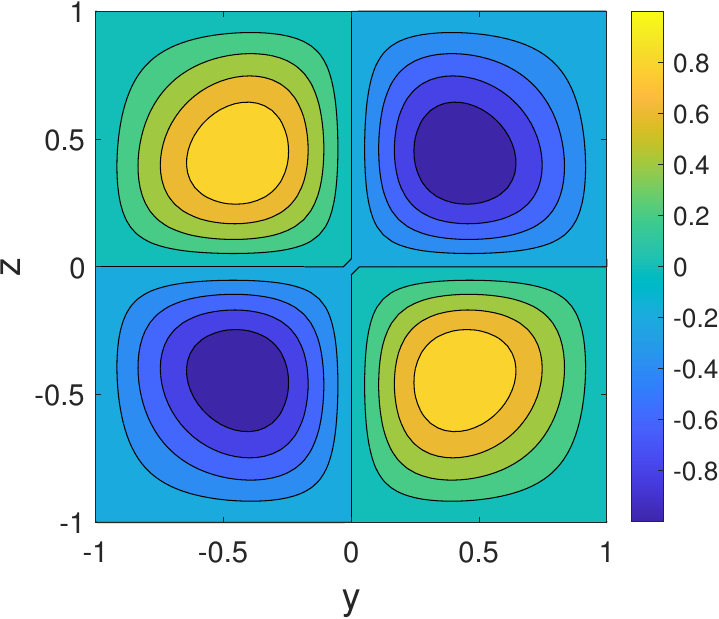}
    \phantomsubcaption
    \label{fig:ductHa0a03_d}
  \end{subfigure}
  \caption{Eigenmodes by ascending $\Rey_E$ from (\subref{fig:ductHa0a03_a}) to (\subref{fig:ductHa0a03_d}) for $\Ha=0$, $\gamma=1$
    and $\alpha=0.3$ visualized by real part of $u_x$. $u_x$ is normalized such that its maximum is equal to unity.
    \mbr{These modes correspond to type I (\subref{fig:ductHa0a03_b}), type II (\subref{fig:ductHa0a03_d}), type III (\subref{fig:ductHa0a03_c}) and type IV (\subref{fig:ductHa0a03_a}).}
    }
  \label{fig:ductHa0a03}
\end{figure}

For a better visualisation of the flow in a cross-section, we chose to display in figure \ref{fig:ductHa0a03} only the real part of the velocity field associated with $u_x$, and chose (before normalisation) the cross-section where the amplitude of \mbr{$\mathrm{Re}(u_x)$} is maximal.
This representation makes the symmetries of the different modes easier to interpret. In particular, figure \ref {fig:ductHa0a03} shows the four least stable modes. We can adopt the nomenclature introduced in the linear stability analyses of \cite{tatsumi1990stability} (for hydrodynamic duct flow) and \cite{priede2010linear} (for the Hunt's flow which admits the same symmetry classification), which is based on listing  whether the symmetry of $u_x$ with respect to the $z$-axis (resp. the $y$-axis) are odd or even. It can be checked that this classification remains unaffected by the presence of the Lorentz force in the governing equations. This gives way to the respective symmetry types I (odd in $y$, even in $z$), II (odd in $y$ and $z$), III (even in $y$ and $z$) and IV (even in $y$/odd in $z$).
%{\color{cyan}MBR: I am confused by our definition of the mode types. We show the streamwise velocity field but Tatsumi \& Yoshimura define these types in terms of the cross-stream components. Uhlmann \& Nagata and Priede et al. give the corresponding symmetries for the streamwise field but they are different from our definition: I (o,e), II (o,o), III (e,e), IV (e,o). So type I and IV are interchanged relative to our definitions. Have I missed something?}
%The modes listed in Figure \ref{fig:ductHa0a03} are, from left to right and from top to bottom, respectively of the type $I,IV,III$ and $II$.
%{\color{red} TB I think I made a mistake when I adapted the definitions from Priede. I thought his magnetic field was along $x$ but it is along $y$ and $z$ is his axial coordinate. Tatsumi use x as streamwise.  Their designation of y and z (to distinguish modes I and IV) is decided by the aspect ratio $A=L_z/L_y\ge 1 $. This cannot be done in our case. To be consistent with Priede we have to pick the magnetic field direction $z$ as second variable and $y$ as first variable to examine the symmetries.  I wrongly used $z$ as first and $y$ as second.
%The correct identifications are $I$ (odd in $y$/even in $z$) and $IV$ (even in $y$/odd in $z$) }
The modes listed in Figure \ref{fig:ductHa0a03} are \mbr{(\subref{fig:ductHa0a03_a}) type IV, (\subref{fig:ductHa0a03_b}) type I, (\subref{fig:ductHa0a03_c}) type III and (\subref{fig:ductHa0a03_d}) type II.}

\begin{figure}
  \centering
  \includegraphics[width=0.5\linewidth]{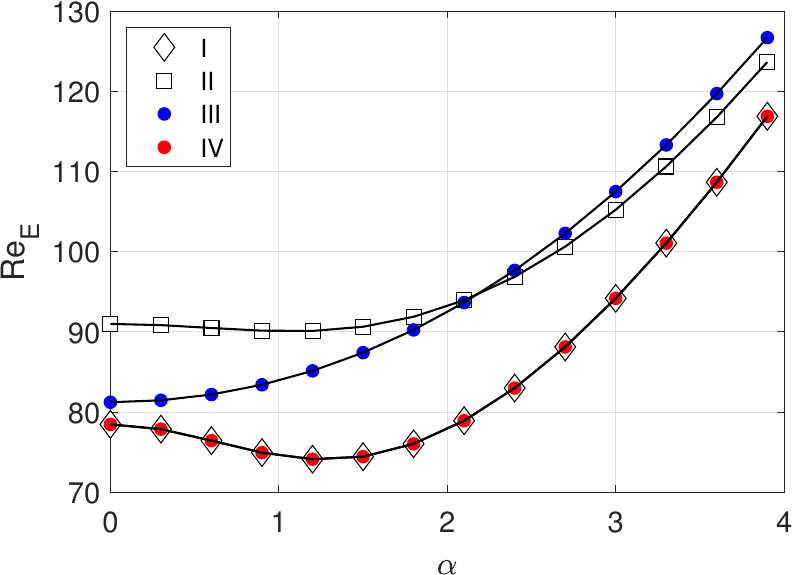}
  \caption{$\Rey_E$ vs.~$ \alpha$ (at optimal wavenumber) for different modes computed for $\Ha=0$,  $\gamma=1$.}
  \label{fig:symmetries}
\end{figure}

\begin{figure}
  \centering
  \begin{subfigure}[b]{0.35\textwidth}
    \setlength{\unitlength}{1.0cm}
    %\fbox{
    \begin{picture}(4.2,2.5)
      \put(-0.1,0.1){\includegraphics[height=2.4cm]{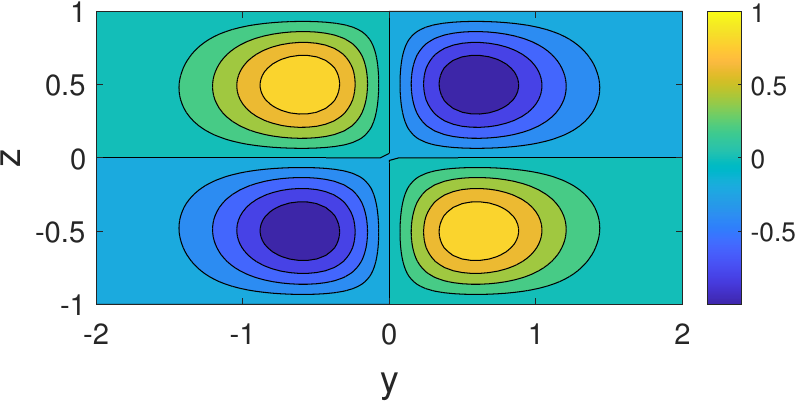}}%
      \put(0.0,0.0){\footnotesize (a)}
    \end{picture}
    %}
    \phantomsubcaption
    \label{fig:ductHa0a0_a}
  \end{subfigure}
  %\hfill
  \begin{subfigure}[b]{0.64\textwidth}
    \setlength{\unitlength}{1.0cm}
    %\fbox{
    \begin{picture}(8.1,2.5)
      \put(-0.1,0.2){\includegraphics[height=2.3cm]{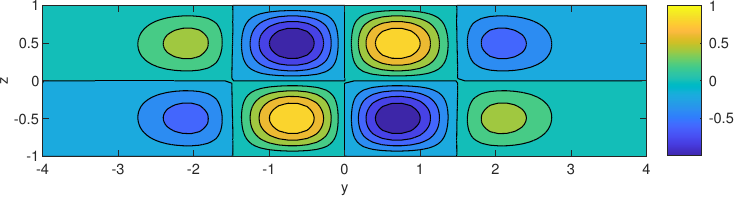}}%
      \put(0.0,0.0){\footnotesize (b)}
    \end{picture}
    %}
    \phantomsubcaption
    \label{fig:ductHa0a0_b}
  \end{subfigure}
  \caption{Lowest eigenmodes \mbr{(type II)} for $\Ha=0$, $\alpha=0$ and (\subref{fig:ductHa0a0_a}) $\gamma=2$, (\subref{fig:ductHa0a0_b}) $\gamma=4$ visualised by $u_x$. $u_x$ is normalized such that its maximum is equal to unity.}
  \label{fig:ductHa0a0}
\end{figure}

The values of $\Rey_E$ corresponding to each of these four modes are, for a more efficient representation, plotted versus their corresponding optimal wavenumber $\alpha$ in Figure \ref{fig:symmetries}. It is clear that the two modes with symmetry I and IV are equivalent for $\gamma=1$ and  the optimal structure for all values of $\alpha$. Modes II and III cross near $\alpha=2$.

For $\Ha=0$, the structure and symmetry type of the lowest mode can change with the aspect ratio $\gamma$. This was already noted in connection with figure \ref{fig:ReEvsgamma_a}. The plots of cross-sections of $u_x$ for $\gamma=2$  and $\gamma=4$ are shown in figure \ref{fig:ductHa0a0_b}. These modes have symmetry II. At $\gamma=4$, the number of structures along the $y$-direction  is twice that for $\gamma=2$.

\subsection{Duct visualisation for $\Ha \neq 0$.}

\begin{figure}
  \centering
  \begin{subfigure}[b]{0.35\textwidth}
    \footnotesize (a) \hspace{-1mm}
    \includegraphics[width=0.9\linewidth]{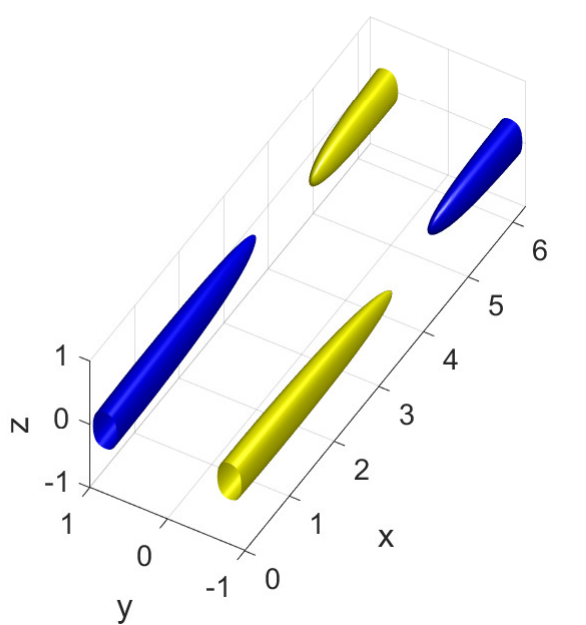}
    \phantomsubcaption
    \label{fig:ductHa203d_a}
  \end{subfigure}
  \begin{subfigure}[b]{0.35\textwidth}
    \footnotesize (b) \hspace{-1mm}
    \includegraphics[width=0.9\linewidth]{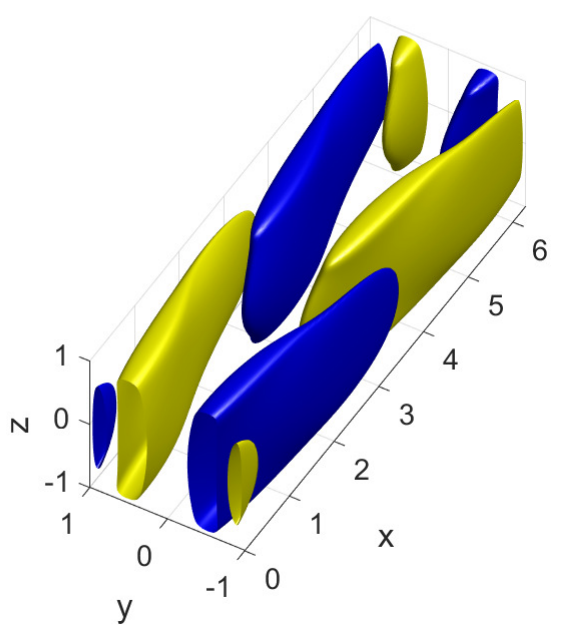}
    \phantomsubcaption
    \label{fig:ductHa203d_b}
  \end{subfigure}\\
  \begin{subfigure}[b]{0.35\textwidth}
    \footnotesize (c) \hspace{-1mm}
    \includegraphics[width=0.9\linewidth]{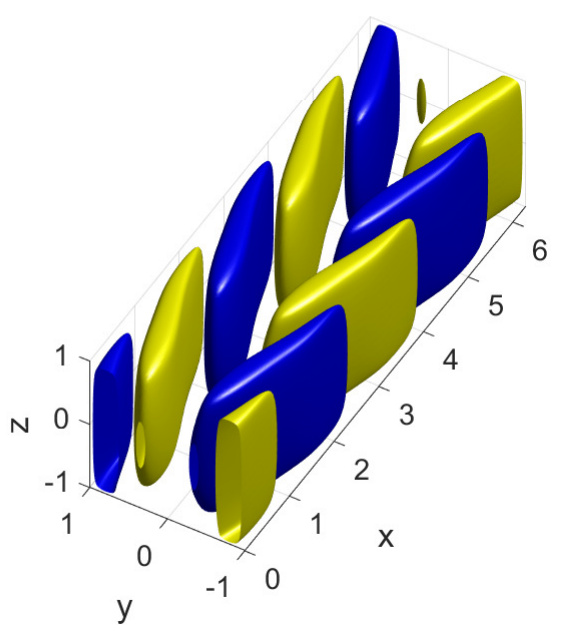}
    \phantomsubcaption
    \label{fig:ductHa203d_c}
  \end{subfigure}
  \begin{subfigure}[b]{0.35\textwidth}
    \footnotesize (d) \hspace{-1mm}
    \includegraphics[width=0.9\linewidth]{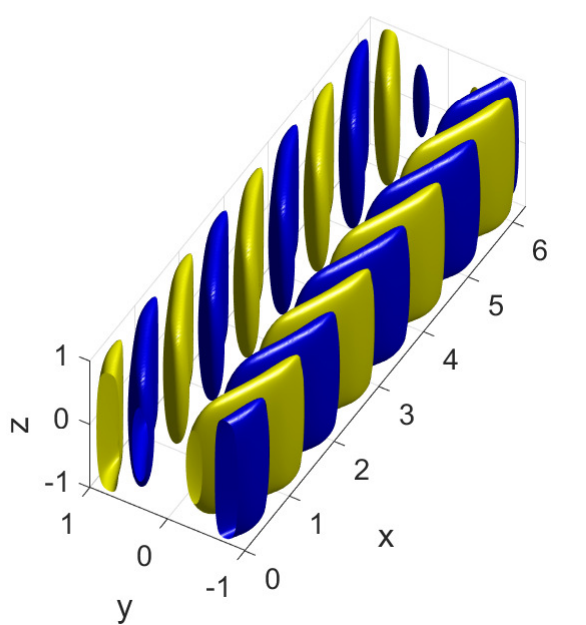}
    \phantomsubcaption
    \label{fig:ductHa203d_d}
  \end{subfigure}
  \caption{Isosurfaces of $u_x$ for the leading eigenmodes at $\Ha=20$, $\gamma=1$ and streamwise wavenumbers (\subref{fig:ductHa203d_a}) $\alpha=1/2$, (\subref{fig:ductHa203d_b}) $\alpha=1$, (\subref{fig:ductHa203d_c}) $\alpha=2$ and (\subref{fig:ductHa203d_d}) $\alpha=4$. \yd{The streamwise period displayed in each case corresponds to the period for $\alpha=1$.}
 % In the case $\alpha=1/2$, only half the  streamwise period is shown.
  }
  \label{fig:ductHa203d}
\end{figure}

\begin{figure}
  \centering
  \begin{subfigure}[b]{0.4\textwidth}
    \footnotesize (a) \hspace{-1mm}
    \includegraphics[width=0.9\linewidth]{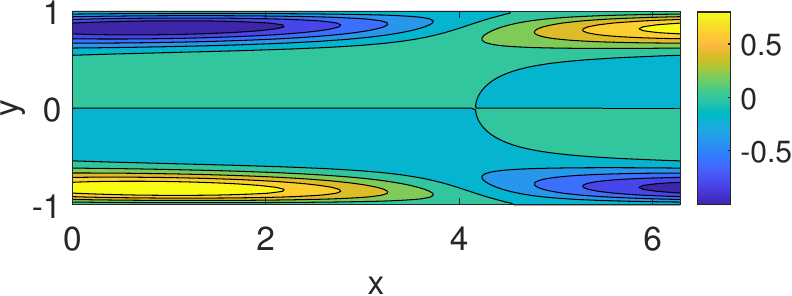}
    \phantomsubcaption
    \label{fig:ductHa203d_slice_a}
  \end{subfigure}
  \begin{subfigure}[b]{0.4\textwidth}
    \footnotesize (b) \hspace{-1mm}
    \includegraphics[width=0.9\linewidth]{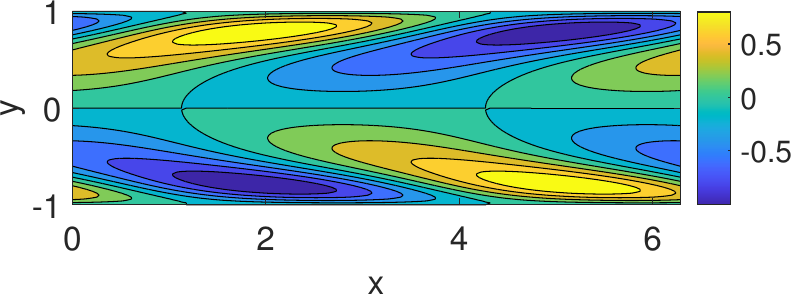}
    \phantomsubcaption
    \label{fig:ductHa203d_slice_b}
  \end{subfigure} \\
  \begin{subfigure}[b]{0.4\textwidth}
    \footnotesize (c) \hspace{-1mm}
    \includegraphics[width=0.9\linewidth]{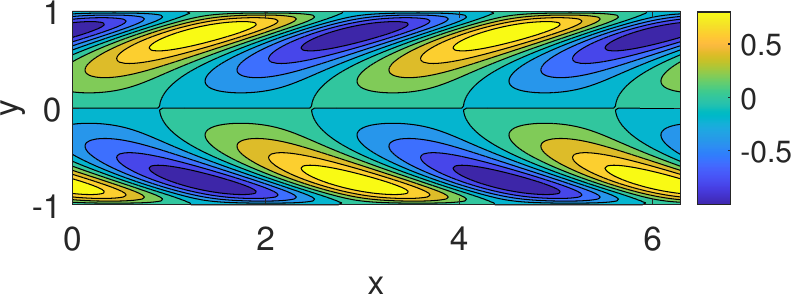}
    \phantomsubcaption
    \label{fig:ductHa203d_slice_c}
  \end{subfigure}
  \begin{subfigure}[b]{0.4\textwidth}
    \footnotesize (d) \hspace{-1mm}
    \includegraphics[width=0.9\linewidth]{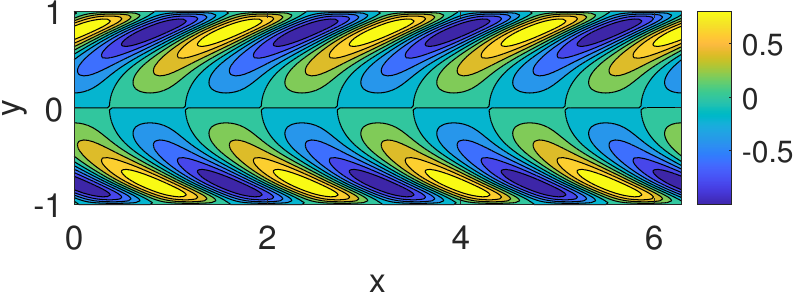}
    \phantomsubcaption
    \label{fig:ductHa203d_slice_d}
  \end{subfigure}
  \caption{
  %Contours of $u_x$ in the plane $z=0$ for the leading eigenmodes at $\Ha=20$, $\gamma=1$ and streamwise wavenumbers (\subref{fig:ductHa203d_slice_a}) $\alpha=1/2$, (\subref{fig:ductHa203d_slice_b}) $\alpha=1$ and (\subref{fig:ductHa203d_slice_c}) $\alpha=2$.
  %In the case $\alpha=1/2$, only half the period is shown.
  \yd{Contours of $u_x$ in the plane $z=0$ for the same parameters as figure \ref{fig:ductHa203d}.}
  }
  \label{fig:ductHa203d_slice}
\end{figure}

\begin{figure}
  \centering
 % \begin{subfigure}[b]{0.32\textwidth}
 %   \footnotesize (a) \hspace{-1mm}
    \includegraphics[width=0.35\linewidth]{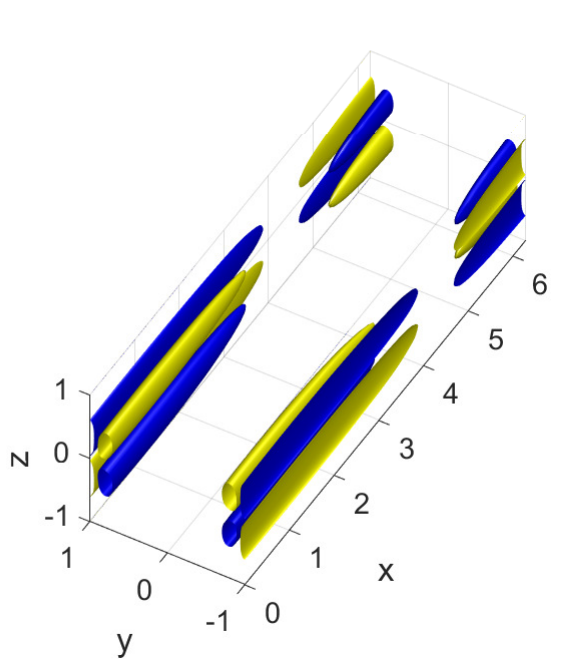}
 %   \phantomsubcaption
 %   \label{fig:ductHa203d_a}
 % \end{subfigure}
 % \begin{subfigure}[b]{0.32\textwidth}
 %   \footnotesize (b) \hspace{-1mm}
 %   \includegraphics[width=0.9\linewidth]{figures/new-omx3deigenmode-1-lx1-beta2-ha20-.eps}
 %   \phantomsubcaption
 %   \label{fig:ductHa203d_b}
 % \end{subfigure}
 % \begin{subfigure}[b]{0.32\textwidth}
 %   \footnotesize (c) \hspace{-1mm}
  %  \includegraphics[width=0.9\linewidth]{figures/new-omx3deigenmode-1-lx1-beta4-ha20-.eps}
  %  \phantomsubcaption
  % \label{fig:ductHa203d_c}
  %\end{subfigure}
  \caption{Isosurfaces of $\omega_x$ for the leading eigenmode at $\Ha=20$, $\gamma=1$ and streamwise wavenumber $\alpha=1/2$. Only half the streamwise period is shown.
  }
  \label{fig:ductHa203d_vort}
\end{figure}

We move next to the visualisation of the eigenmodes of the eigenproblem \eqref{eq:energystability2} in the MHD case.
Figure \ref{fig:ductHa203d} shows three-dimensional visualisations of the least stable modes found in square duct ($\gamma=1$) at $\Ha=20$ \yd{for $\alpha=1/2,1,2$ and $4$}. Again, long modes with small or vanishing $\alpha$ are associated at higher $\Rey$ with the lift-up mechanism while the Orr mechanism is present in shorter-wavelength structures. A clear difference is that the structures are more localized near the side walls since the central part of the velocity distribution has low shear.
\tb{As for $\Ha=0$ the streamwise velocity has also been plotted in the $xy$ plane where variations along $y$ are important (see figure \ref{fig:ductHa203d_slice}), whereas in the $xz$ plane the perturbations show a more uniform dependence on $z$.
The tilting against the shear, clearly visible in the Shercliff layers, is noted for $\alpha=1,2$
as well as for $\alpha=4$ where the lowest value of $\Rey_E$ is achieved for $\gamma=1$.}
 For smaller $\alpha=1/2$, the streaky perturbations are found in the Shercliff layers only and are less elongated along $z$ than for the higher $\alpha$ values. \yd{Isosurfaces of the corresponding streamwise vorticity $\omega_x$ are shown in figure \ref{fig:ductHa203d_vort}. Their position inside the Shercliff layers is consistent with the emergence of streamwise streaks via the lift-up mechanism.}

%\begin{figure}
%  \centering
%%  \begin{subfigure}[b]{0.32\textwidth}
%    \footnotesize (a) \hspace{-1mm}
%    \includegraphics[width=0.9\linewidth]{figures/ux-yconst-1-lx1-beta1-ha20-.eps}
%    \phantomsubcaption
%    \label{fig:ductHa203d_a}
%  \end{subfigure}
%  \begin{subfigure}[b]{0.32\textwidth}
%    \footnotesize (b) \hspace{-1mm}
%    \includegraphics[width=0.9\linewidth]{figures/ux-yconst-1-lx1-beta2-ha20-.eps}
%    \phantomsubcaption
%    \label{fig:ductHa203d_b}
%  \end{subfigure}
%  \begin{subfigure}[b]{0.32\textwidth}
%    \footnotesize (c) \hspace{-1mm}
%    \includegraphics[width=0.9\linewidth]{figures/ux-yconst-1-lx1-beta4-ha20-.eps}
%    \phantomsubcaption
%    \label{fig:ductHa203d_c}
%  \end{subfigure}
%  \caption{Contours of $u_x$ in the plane $y=1/2$ for the leading eigenmodes at $\Ha=20$, $\gamma=1$ and streamwise wavenumbers (\subref{fig:ductHa203d_a}) $\alpha=1/2$, (\subref{fig:ductHa203d_b}) $\alpha=1$ and (\subref{fig:ductHa203d_c}) $\alpha=2$.
%  In the case $\alpha=1/2$, only half the period is shown.
 % }
%  \label{fig:ductHa203d}
%\end{figure}

\begin{figure}
  \centering
  \includegraphics[width=0.5\linewidth]{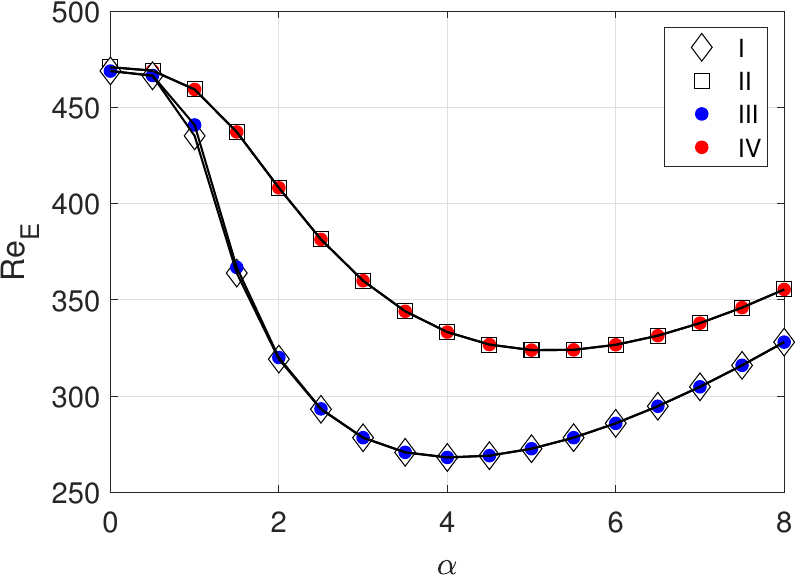}
  \caption{$\Rey_E$ vs.~$ \alpha$ for different modes computed for $\Ha=20$, $\gamma=1$.}
  \label{fig:symmetries20}
\end{figure}

\begin{figure}
  \centering
  \begin{subfigure}[b]{0.49\textwidth}
    \footnotesize (a) \hspace{-1mm}
    \includegraphics[width=0.9\linewidth]{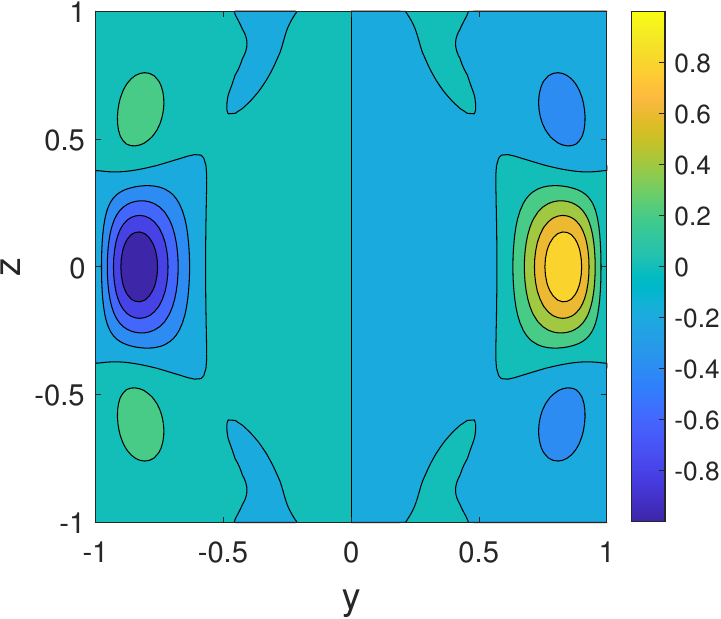}
    \phantomsubcaption
    \label{fig:ductHa20a12_a}
  \end{subfigure}
  \hfill
  \begin{subfigure}[b]{0.49\textwidth}
    \footnotesize (b) \hspace{-1mm}
    \includegraphics[width=0.9\linewidth]{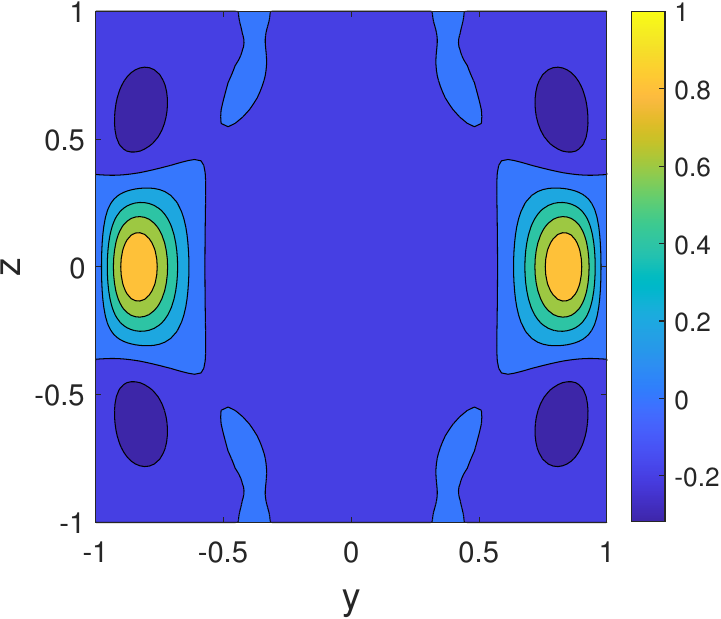}
    \phantomsubcaption
    \label{fig:ductHa20a12_b}
  \end{subfigure}\\
  \begin{subfigure}[b]{0.49\textwidth}
    \footnotesize (c) \hspace{-1mm}
    \includegraphics[width=0.9\linewidth]{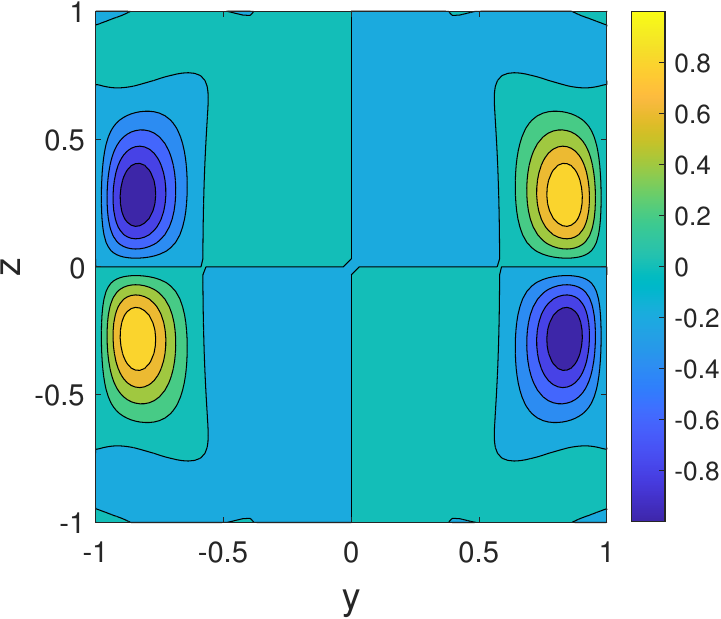}
    \phantomsubcaption
    \label{fig:ductHa20a12_c}
  \end{subfigure}
  \hfill
  \begin{subfigure}[b]{0.49\textwidth}
    \footnotesize (d) \hspace{-1mm}
    \includegraphics[width=0.9\linewidth]{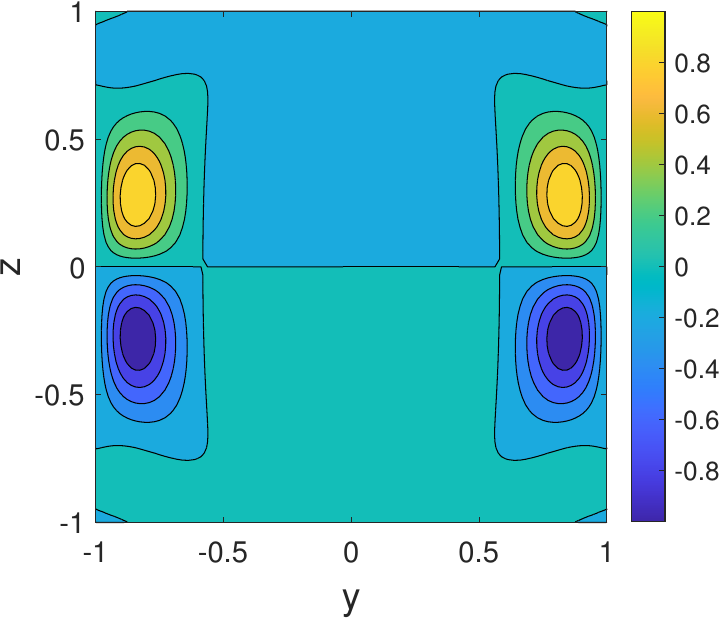}
    \phantomsubcaption
    \label{fig:ductHa20a12_d}
  \end{subfigure}
  \caption{Eigenmodes by ascending $\Rey_E$ from (\subref{fig:ductHa20a12_a}) to (\subref{fig:ductHa20a12_d}) for $\Ha=20$, $\gamma=1$ and $\alpha=1/2$ visualized by real part of $u_x$. $u_x$ is normalized such that its maximum is equal to unity.
  \mbr{These modes correspond to type I (\subref{fig:ductHa20a12_a}), type II (\subref{fig:ductHa20a12_c}), type III (\subref{fig:ductHa20a12_b}) and type IV (\subref{fig:ductHa20a12_d}).}
  }
  \label{fig:ductHa20a12}
\end{figure}

\begin{figure}
  \centering
  \begin{subfigure}[b]{0.49\textwidth}
    \footnotesize (a) \hspace{-1mm}
    \includegraphics[width=0.9\linewidth]{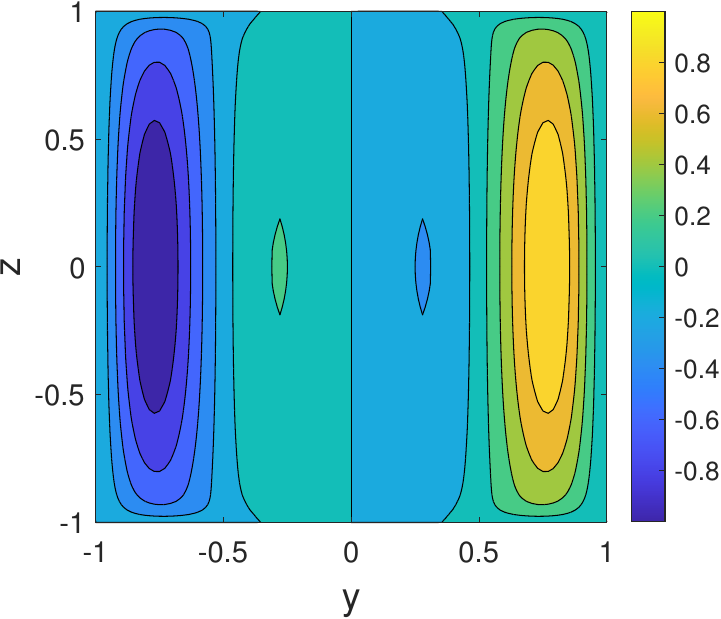}
    \phantomsubcaption
    \label{fig:ductHa20a2_a}
  \end{subfigure}
  \hfill
  \begin{subfigure}[b]{0.49\textwidth}
    \footnotesize (b) \hspace{-1mm}
    \includegraphics[width=0.9\linewidth]{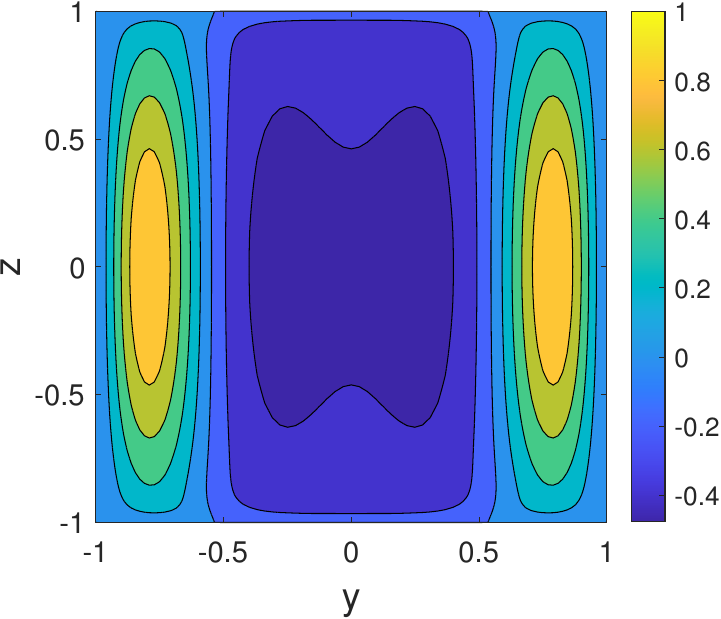}
    \phantomsubcaption
    \label{fig:ductHa20a2_b}
  \end{subfigure}\\
  \begin{subfigure}[b]{0.49\textwidth}
    \footnotesize (c) \hspace{-1mm}
    \includegraphics[width=0.9\linewidth]{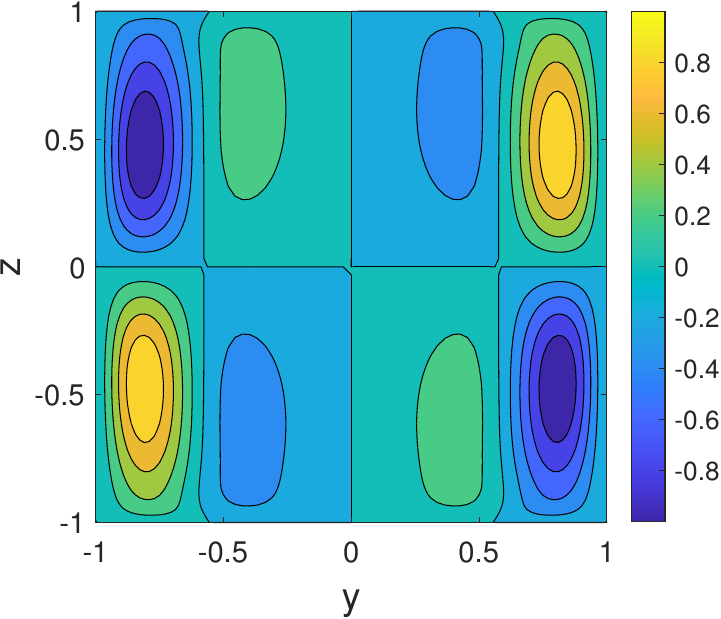}
    \phantomsubcaption
    \label{fig:ductHa20a2_c}
  \end{subfigure}
  \hfill
  \begin{subfigure}[b]{0.49\textwidth}
    \footnotesize (d) \hspace{-1mm}
    \includegraphics[width=0.9\linewidth]{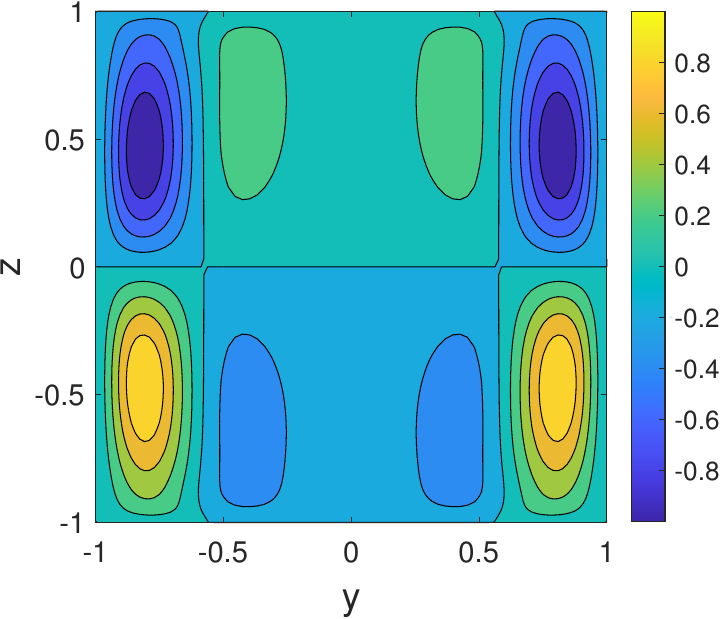}
    \phantomsubcaption
    \label{fig:ductHa20a2_d}
  \end{subfigure}
  \caption{Eigenmodes by ascending $\Rey_E$ from (\subref{fig:ductHa20a2_a}) to (\subref{fig:ductHa20a2_d}) for $\Ha=20$, $\gamma=1$ and $\alpha=2$ visualized by real part of $u_x$. $u_x$ is normalized such that its maximum is equal to unity.
  \mbr{These modes correspond to type I (\subref{fig:ductHa20a2_a}), type II (\subref{fig:ductHa20a2_c}), type III (\subref{fig:ductHa20a2_b}) and type IV (\subref{fig:ductHa20a2_d}).}
  }
  \label{fig:ductHa20a2}
\end{figure}

\begin{figure}
  \centering
  \begin{subfigure}[b]{0.49\textwidth}
    \footnotesize (a) \hspace{-1mm}
    \includegraphics[width=0.9\linewidth]{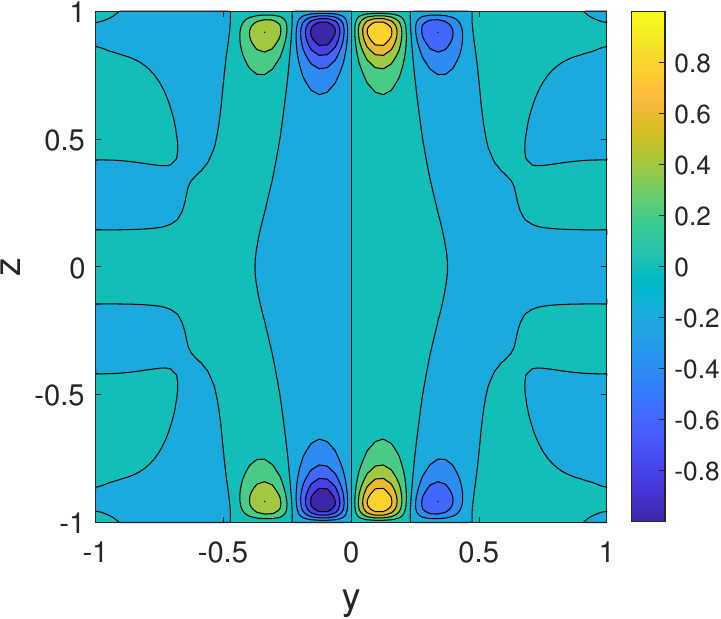}
    \phantomsubcaption
    \label{fig:ductHa20higher_a}
  \end{subfigure}
  \hfill
  \begin{subfigure}[b]{0.49\textwidth}
    \footnotesize (b) \hspace{-1mm}
    \includegraphics[width=0.9\linewidth]{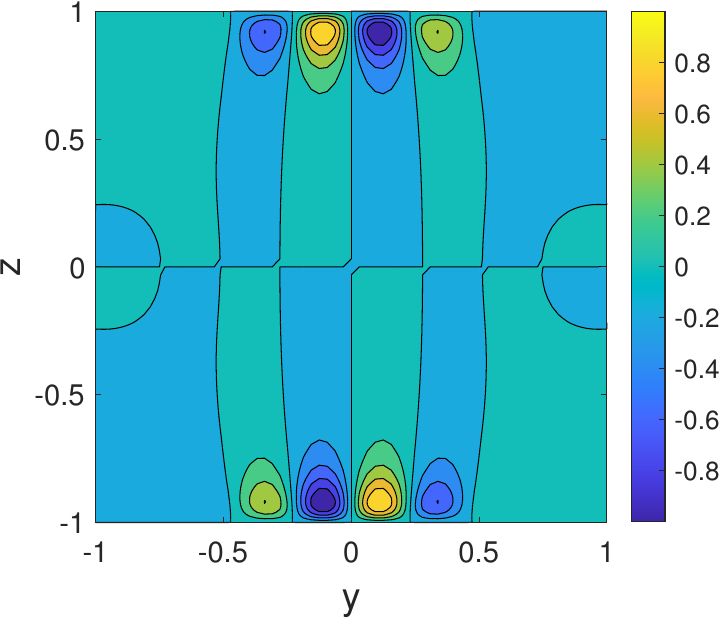}
    \phantomsubcaption
    \label{fig:ductHa20higher_b}
  \end{subfigure}\\
  \begin{subfigure}[b]{0.49\textwidth}
    \footnotesize (c) \hspace{-1mm}
    \includegraphics[width=0.9\linewidth]{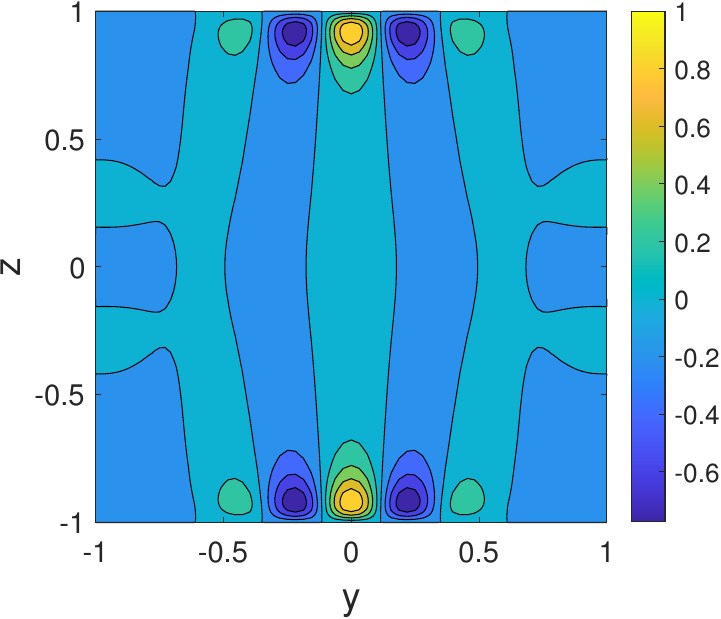}
    \phantomsubcaption
    \label{fig:ductHa20higher_c}
  \end{subfigure}
  \hfill
  \begin{subfigure}[b]{0.49\textwidth}
    \footnotesize (d) \hspace{-1mm}
    \includegraphics[width=0.9\linewidth]{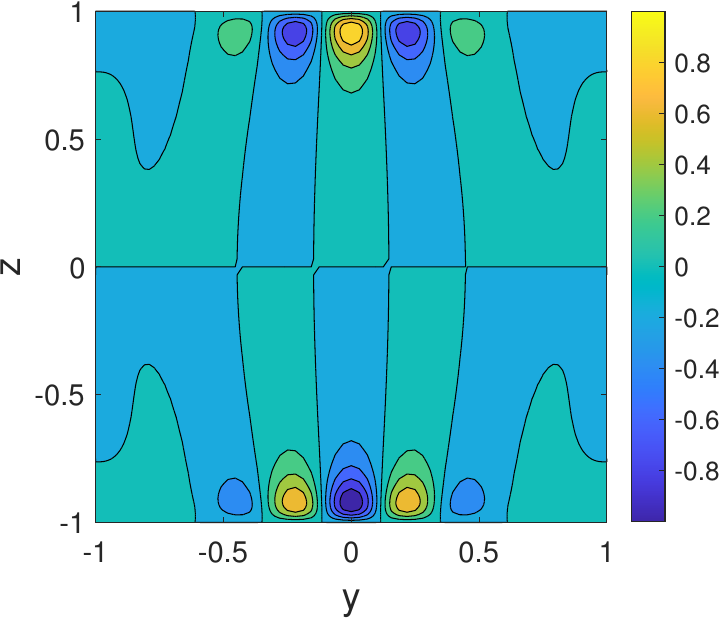}
    \phantomsubcaption
    \label{fig:ductHa20higher_d}
  \end{subfigure}
  \caption{Selection of eigenmodes for $\Ha=20$, $\gamma=1$ and $\alpha=0$ visualized by $u_x$.
  (\subref{fig:ductHa20higher_a}) mode 9 with $\Rey_E=530.16$,
  (\subref{fig:ductHa20higher_b}) mode 10 with $\Rey_E=530.27$,
  (\subref{fig:ductHa20higher_c}) mode 11 with $\Rey_E=530.32$,
  (\subref{fig:ductHa20higher_d}) mode 12 with $\Rey_E=530.33$.
  $u_x$ is normalized such that its maximum is equal to unity.
  }
  \label{fig:ductHa20higher}
\end{figure}

For the least stable modes found for $\gamma=1$, this leads to the figure \ref{fig:symmetries20} where $\Rey_E$ for a given mode is plotted versus the corresponding wavenumber $\alpha$.
A clear departure from the $\Ha=0$ case is observed. This is mainly due to the fact that $\alpha=0$ perturbations are much more damped than non-zero ones. In particular, the value for $\Rey_E$ starts from 470 for $\alpha=0$ (to be compared to a value of less than 100 for $\Ha=0$), then it decreases rapidly to values closer to 250-300 in the range $3\le \alpha \le 4$, only to start rising beyond that. Almost identical values of $\Rey_E$ are obtained for the symmetry types III and I as well as IV and II, \ie with either even or odd symmetry in $z$. The lateral symmetry, therefore, does not matter except for small $\alpha$. Cross-sections of the real part of the $u_x$ modes are again shown in figures \ref{fig:ductHa20a12}, \ref{fig:ductHa20a2} and \ref{fig:ductHa20higher}.
For both $\alpha=1/2$ (figure \ref{fig:ductHa20a12}) and $\alpha=2$ (figure \ref{fig:ductHa20a2}), the perturbations are clearly located in the Shercliff layers with symmetries \mbr{type I (\ref{fig:ductHa20a12_a} and \ref{fig:ductHa20a2_a}), III (\ref{fig:ductHa20a12_b} and \ref{fig:ductHa20a2_b}), II (\ref{fig:ductHa20a12_c} and \ref{fig:ductHa20a2_c}) and IV (\ref{fig:ductHa20a12_d} and \ref{fig:ductHa20a2_d}).}
%For both $\alpha=1/2$ (figure \ref{fig:ductHa20a12}) and $\alpha=2$ (figure \ref{fig:ductHa20a2}), the perturbations are clearly located in the Shercliff layers with symmetries $IV$, $III$, $II$ and $I$ (left to right and top to bottom).
It is again apparent that the structures become more elongated along $z$ for the higher $\alpha$ in figure \ref{fig:ductHa20a2}.

\yd{Not all eigenmodes found in this study are located in the Shercliff layers,
although those which minimise the value of $Re_E$ generally are. The existence of modes localized inside the Hartmann layers, already mentioned in \S\ref{sec:alphazero}, is demonstrated in figure \ref{fig:ductHa20higher}. These different modes are of type I, II, III and IV, respectively (modes 9-12). The corresponding values of $Re_E$ are all close to 530, which is roughly twice the global minimising value of $Re_E$ for these parameters.}
Owing to the wide separation between the Hartmann layers, the symmetry with respect to $z$ \yd{has little influence on the value of $Re_E$}. The symmetry with respect to $y$ hardly affects the eigenvalues either.
%\textcolor{cyan}{\sout{although the even symmetry seems to be marginally preferred and provides somewhat larger structures in the Hartmann layers} [MBR: I DO NOT SEE THIS!]}.

\section{Summary and conclusions}
\label{sec:conclusions}

In this computational study, energy stability theory was applied to the case of hydrodynamic and magnetohydrodynamic duct flow, in the situation where the flow is electrically conducting, the walls electrically insulating, and the applied magnetic field is transverse. The duct is assumed periodic in the streamwise direction. The values of the energy Reynolds number $\Rey_E$ were reported in a parametric study accounting for variable streamwise wavenumber $\alpha$, variable cross-sectional aspect ratio $\gamma$ and variable Hartmann number (which is proportional the intensity of the applied magnetic field). By going to the $\gamma \ll 1$ limit, the results for spanwise-periodic channel flow were recovered. For $\gamma \gg 1$, they also match with the quasi-two-dimensional (Q2D) computations performed in \cite{potherat2007quasi} for short streamwise wavelengths. \yd{The related perturbations are found along the sidewalls (\eg inside the Shercliff layers).} The special case of streamwise-independent perturbations shows a different scaling than in the Q2D case, which agrees with the arguments developed in \cite{krasnov2010optimal}.
The visualisation of the critical structures, interpreted as precursors of the popular linear optimal perturbations (LOPs), allows one to identify the mechanisms at play at $\Rey_E$, namely the Orr and the lift-up mechanism. All these optimal structures are found to be robustly located in the Shercliff layer. This result is nicely consistent with the localisation of the linear optimal modes reported in \cite{cassells2019three}. Note that the Shercliff layer is also the part of the cross-section where turbulence remains last in the spatio-temporally intermittent regime, as $\Ha$ is increased \citep{krasnov2013patterned}. This robust property highly suggests, by extrapolating the system to yet higher $\Rey$ and to the fully nonlinear regime, that transition to turbulence in this geometry starts \textit{generically} with the destabilisation of the Shercliff layer. Such conclusions need to be supported by nonlinear computations which are currently underway \citep{brynjell2022edge}.

%\backsection[Supplementary data]{\label{SupMat}Supplementary material and movies are available at \\https://doi.org/10.1017/jfm.2019...}

\backsection[Acknowledgements]{The authors thank the Computing Center of Technische Universit{\"a}t Ilmenau for providing the software and computing time on its computer cluster.
}

\backsection[Funding]{
  The authors acknowledge financial suppport from the Deutsche Forschungsgemeinschaft (project number 470628784).
}

\backsection[Declaration of interests]{The authors report no conflict of interest.}

%\backsection[Data availability statement]{The data that support the findings of this study are openly available in [repository name] at http://doi.org/[doi], reference number [reference number]. See JFM's \href{https://www.cambridge.org/core/journals/journal-of-fluid-mechanics/information/journal-policies/research-transparency}{research transparency policy} for more information}

\backsection[Author ORCIDs]{
  Thomas Boeck, https://orcid.org/0000-0002-0814-7432; Mattias Brynjell-Rahkola, https://orcid.org/0000-0001-9446-7477; Yohann Duguet, https://orcid.org/0000-0001-6258-0475}

%\backsection[Author contributions]{Authors may include details of the contributions made by each author to the manuscript'}
\appendix
\section{Construction of matrices in the eigenvalue problem (\ref{eq:generalizedevp})}
\label{sec:appendix_matrices}
In order to describe how the matrices are assembled
we focus on the case with non-zero streamwise wavenumber, \ie on the discretization of equations (\ref{eq:estab-vectorsf1}-\ref{eq:estab-vectorsf5}).
The case of zero streamwise wavenumber can be treated similarly except that the relevant equations  are (\ref{eq:estab-scalarsf1}-\ref{eq:estab-scalarsf4}). They are discretised by the same approach, which  should be apparent from the following discussion.

Since the vorticity components $\hat{\omega}_y$, $\hat{\omega}_z$ are given by the Laplacians of $\hat{\psi}_y$ and $\hat{\psi}_z$, $\hat{\phi}$ (\cf equations (\ref{eq:estab-vectorsf3},\ref{eq:estab-vectorsf4},\ref{eq:estab-vectorsf5})), we chose a larger set of basis functions for these quantities. We use $(N_y+1)\times (N_z+1)$ Chebyshev polynomials for the two vorticity components and $(N_y+3)\times (N_z+3)$ Chebyshev polynomials for the vector streamfunction components and the electric potential.

We write these two sets $\{g_m\}$ and  $\{h_m\}$ using a combined single index rather than  double indices. The definitions are
\begin{equation}\label{eq:app-gm}
g_{k+1+l(N_y+3)}(y,z)=T_k(y/L_y) T_l(z/L_z), \quad 0\le k\le N_y+2,\quad 0\le l\le N_z+2
\end{equation}
%for the representation of the vector streamfunction components and the electric potential and
and
\begin{equation}\label{eq:app-hm}
h_{k+1+l(N_y+1)}(y,z)=T_k(y/L_y) T_l(z/L_z), \quad 0\le k\le N_y,\quad 0\le l\le N_z.
\end{equation}
The expansion for $\hat{\omega}_y$ then reads
\begin{equation}\label{eq:app-expansion-omegay}
\hat{\omega}_y=\sum_{m=1}^{N_{\max,1}} \Omega^{(y)}_m \, h_m(y,z),\quad N_{\max,1}= (N_y+1)(N_z+1).
\end{equation}
Analogously,  the expansion for $\hat{\psi}_y$ is
\begin{equation}\label{eq:app-expansionp-siy}
\hat{\psi}_y=\sum_{m=1}^{N_{\max,2}} \Psi^{(y)}_m \, g_m(y,z),\quad N_{\max,2}= (N_y+3)(N_z+3).
\end{equation}
For the discretisation of the Poisson equation (\ref{eq:estab-vectorsf3}) we demand that the partial differential equation holds at the interior collocation points
\begin{equation}\label{eq:app-collocationpoints-large}
(y_k, z_l) = \left( L_y\cos\left(\frac{k\pi}{N_y+2}\right),L_z\cos\left(\frac{l\pi}{N_z+2}\right)\right).
%,  1\le k\le N_y+1,\quad 1\le l\le N_z+1
\end{equation}
To wit,
%{\color{cyan}MBR: What does 'wit' mean?}
\begin{equation}\label{eq:app-collocation-psiy-omegay-interior}
\sum_{m=1}^{N_{\max,2}}\Psi^{(y)}_m \underbrace{\left(\left.\frac{\partial^2 g_m}{\partial y^2}\right|_{(y_k,z_l)}+ \left.\frac{\partial^2 g_m}{\partial z^2}\right|_{(y_k,z_l)}  - \alpha^2 g_m(y_l,z_k)\right)}_{=\mathsf{Q}_{i,m}}= \sum_{m=1}^{N_{\max,1}} \Omega^{(y)}_m\, \underbrace{h_m(y_k,z_l)}_{=\mathsf{R}_{i,m}},
\end{equation}
where $i=k+(l-1)(N_y+1)$,  $1\le k\le N_y+1$, $ 1\le l\le N_z+1$. For $i\le N_{\max,1}$, the matrix elements $\mathsf{R}_{i,m}$ with $m>N_{\max,1}$ are set to zero.

The remaining equations for  $N_{\max,1} \le i \le N_{\max,2}$ are obtained from the boundary conditions, which are imposed on the collocation points on the boundary. On the line $z=L_z$ the value of $\hat{\psi_y}$ is prescribed, \eg
\begin{equation}\label{eq:app-collocation-psiy-omegay-boundary}
\sum_{m=1}^{N_{\max,2}}\Psi^{(y)}_m \underbrace{ g_m(y_l, L_z )}_{=\mathsf{Q}_{i,m}}= v_i,
\end{equation}
where $v_i$ denotes the boundary value and $i=N_{\max,1}+l+1$ with $ 0\le l\le N_y+2$.  This equation implies that $\mathsf{R}$ has only diagonal entries for $i>N_{\max,1}$, which can be set to unity.
The remaining boundaries $z=-L_z$, $y=\pm L_y$ are treated in a similar way.  The set of boundary values $v_i$ complements the set of known values $\Omega^{(y)}_m$. For simplicity we define them as $s_i=\Omega^{(y)}_i$ for $i\le N_{\max,1}$ and $s_i=v_i=0$ for  $N_{\max,1} < i \le N_{\max,2}$. In summary, one obtains a linear system
\begin{equation}\label{eq:app-linearsystem-psiy-omegay}
\sum_{m=1}^{N_{\max,2}} \mathsf{Q}_{i,m}\Psi^{(y)}_m=\sum_{m=1}^{N_{\max,2}} \mathsf{R}_{i,m} s_m
\end{equation}
with full-rank matrices $\mathsf{Q}$ and $\mathsf{R}$. One can therefore compute  $\mathsf{C}^{(y)}=\mathsf{Q}^{-1}\mathsf{R}$ and represent the vector $\Psi^{(y)}_m$ by
\begin{equation}
\Psi^{(y)}_i = \sum_{m=1}^{N_{\max,2}} \mathsf{C}^{(y)}_{i,m}\, s_m.
\end{equation}
Since the boundary values $v_i$ are all zero, this reduces to
\begin{equation}\label{eq:app-psiy-through-omegay}
\Psi^{(y)}_i = \sum_{m=1}^{N_{\max,1}} \mathsf{C}^{(y)}_{i,m}\, \Omega^{(y)}_m, \quad 1\le i \le N_{\max,2}.
\end{equation}
The expansion coefficients $\Psi^{(z)}_i$ and $\Phi_i$ are obtained in an analogous way from eqs.~(\ref{eq:estab-vectorsf4}) and (\ref{eq:estab-vectorsf5}). Since boundary values are again zero, we have
\begin{equation}\label{eq:app-psiz-through-omegaz}
\Psi^{(z)}_i = \sum_{m=1}^{N_{\max,1}} \mathsf{C}^{(z)}_{i,m}\, \Omega^{(z)}_m, \quad 1\le i \le N_{\max,2}
\end{equation}
and
\begin{equation}\label{eq:app-phi-through-omegaz}
\Phi_i = \sum_{m=1}^{N_{\max,1}} \mathsf{C}^{(\phi)}_{i,m}\, \Omega^{(z)}_m, \quad 1\le i \le N_{\max,2}.
\end{equation}

For the discretization of eq. (\ref{eq:estab-vectorsf1},\ref{eq:estab-vectorsf2})  we use the smaller set of $(N_y-1)\times (N_z-1)$ interior collocation points corresponding to the set $\{h_m\}$, namely
\begin{equation}\label{eq:app-collocationpoints-small}
(y_k, z_l) = \left( L_y\cos\left(\frac{k\pi}{N_y}\right),L_z\cos\left(\frac{l\pi}{N_z}\right)\right).
%,  1\le k\le N_y+1,\quad 1\le l\le N_z+1
\end{equation}
Most elements of matrices $\bm{\mathsf{A}}$ and $\bm{\mathsf{B}}$ are obtained by demanding that the equations  (\ref{eq:estab-vectorsf1},\ref{eq:estab-vectorsf2})  hold  at these collocation points. The remaining ones are obtained from the boundary conditions applied to the collocation points on the boundary.

It is straightforward but tedious to evaluate the contributions from the different terms appearing in the equations  (\ref{eq:estab-vectorsf1},\ref{eq:estab-vectorsf2}). The Laplacian on the left hand side of eq.  (\ref{eq:estab-vectorsf1})  yields
\begin{equation}\label{eq:app-collocation-delta-omegay-interior}
\sum_{m=1}^{N_{\max,1}}\Omega^{(y)}_m \left(\left.\frac{\partial^2 h_m}{\partial y^2}\right|_{(y_k,z_l)}+ \left.\frac{\partial^2 h_m}{\partial z^2}\right|_{(y_k,z_l)}  - \alpha^2 h_m(y_l,z_k)\right),
\end{equation}
where the term multiplying $\Omega^{(y)}_m$ adds to the entry $\bm{\mathsf{A}}_{i,m}$ where
$i=k+(l-1)(N_y-1)$. The contribution from the second term is
\begin{equation}\label{eq:app-collocation-dxdz-phi-interior}
-\Ha^2 \sum_{m=1}^{N_{\max,2}}\Phi_m \left.\frac{\partial^2 g_m}{\partial y\partial z}\right|_{(y_k,z_l)}= \sum_{n=1}^{N_{\max,1}}  \Omega^{(z)}_n \left(\sum_{m=1}^{N_{\max,2}} -\Ha^2 \mathsf{C}^{(\phi)}_{m,n}\, \left.\frac{\partial^2 g_m}{\partial y\partial z}\right|_{(y_k,z_l)}\right)
\end{equation}
We see that both $\Omega^{(y)}_m$ and  $\Omega^{(z)}_m$ appear in the discretisation of eq. (\ref{eq:estab-vectorsf1}). It is therefore necessary to
%The underlined contribution comes from the vertical vorticity, i.e. one has to
 combine  $\Omega^{(y)}_m$ and $\Omega^{(z)}_m$ into a single vector $\bm{Y}$ of size $2N_{\max,1}$ with
$Y_m=\Omega^{(y)}_m$  and $Y_{m+N_{\max,1}}=\Omega^{(z)}_m$ for $1 \le m \le N_{\max,1}$. The matrices $\bm{\mathsf{A}}$ and $\bm{\mathsf{B}}$ are then also of size $2N_{\max,1}\times 2N_{\max,1}$. The term multiplying $\Omega^{(z)}_n$ %The underlined term
on the right hand side of eq.~(\ref{eq:app-collocation-dxdz-phi-interior}) adds  to the matrix element $\bm{\mathsf{A}}_{i,n+N_{\max,1}}$. The treatment of the other terms contributing to $\bm{\mathsf{A}}$ and $\bm{\mathsf{B}}$ at the interior collocation points (\ref{eq:app-collocationpoints-small}) is analogous.

The boundary conditions on $\hat{\omega}_y$ and $\hat{\omega}_z$ are implemented on the remaining rows of matrix $\bm{\mathsf{A}}$, i.e. for index values $2 (N_y-1)(N_z-1) < i\le 2N_{\max,1}$ as explained above for the computation of $\hat{\psi}_y$ from $\hat{\omega}_y$. Since the boundary conditions do not contain the eigenvalue $\Rey$, the corresponding rows of matrix $\bm{\mathsf{B}}$ are filled with zeros.

\bibliographystyle{jfm}

%\bibliography{literature}

\end{document}